\RequirePackage{fix-cm}
\documentclass[smallextended]{svjour3}
\usepackage[numbers,sort&compress]{natbib}
\usepackage[colorlinks,urlcolor=blue,citecolor=blue,linkcolor=blue]{hyperref}
\smartqed
\usepackage{amssymb}
\usepackage{amsmath}
\usepackage{graphicx}
\usepackage{dcolumn}
\usepackage{hyperref}
\usepackage{color}
\usepackage{units}
\usepackage[dvipsnames]{xcolor} 
\usepackage{aas_macros}

\usepackage[normalem]{ulem} 

\newcommand{\done}[1]{\textcolor{green}{[done]}}
\newcommand{\spa}[1]{School of Physics and Astronomy, Monash University, Vic 3800, Australia}
\newcommand{\ozgrav}[1]{OzGrav: The ARC Centre of Excellence for Gravitational Wave Discovery, Clayton VIC 3800, Australia}
\newcommand{\citeg}[1]{\citep[e.g.,][]{#1}}

\begin{document}

\title{The evolution of binary neutron star post-merger remnants: a review
}
\titlerunning{Observational signatures of merger remnants}
\author{Nikhil Sarin and Paul D.~Lasky}
\authorrunning{N.Sarin and P.D.~Lasky}
\institute{N.Sarin and P.D.~Lasky\at\spa\\
\ozgrav \\ \\
\email{nikhil.sarin@monash.edu; paul.lasky@monash.edu}}
\date{Received: date / Accepted: date}
\maketitle

\begin{abstract}
Two neutron stars merge somewhere in the Universe approximately every 10 to 100 seconds, creating violent explosions potentially observable in gravitational waves and across the electromagnetic spectrum. 
The transformative coincident gravitational-wave and electromagnetic observations of the binary neutron star merger GW170817 gave invaluable insights into these cataclysmic collisions, probing bulk nuclear matter at supranuclear densities, the jet structure of gamma-ray bursts, the speed of gravity, and the cosmological evolution of the local Universe, among other things.
Despite the wealth of information, it is still unclear when the remnant of GW170817 collapsed to form a black hole. 
Evidence from other short gamma-ray bursts indicates a large fraction of mergers may form long-lived neutron stars.  
We review what is known observationally and theoretically about binary neutron star post-merger remnants.  
From a theoretical perspective, we review our understanding of the evolution of short- and long-lived merger remnants, including fluid, magnetic-field, and temperature evolution.  
These considerations impact prospects of detection of gravitational waves from either short- or long-lived neutron star remnants which potentially allows for new probes into the hot nuclear equation of state in conditions inaccessible in terrestrial experiments.
We also review prospects for determining post-merger physics from current and future electromagnetic observations, including kilonovae and late-time x-ray and radio afterglow observations.
\keywords{Binary neutron star mergers \and remnants \and gravitational-waves \and gamma-ray bursts \and kilonovae}
\end{abstract}

\section{Introduction}
The coincident gravitational-wave and electromagnetic observations of binary neutron star merger GW170817/GRB170817A~\cite{abbott17_gw170817_detection,abbott17_gw170817_multimessenger,abbott17_gw170817_gwgrb} was a watershed moment, signaling the beginning of a new field of multi-messenger gravitational-wave astronomy.  Gravitational-wave emission from the inspiral phase was detected by the Advanced LIGO~\cite{LIGO} and Advanced Virgo~\cite{Virgo} interferometers~\cite{abbott17_gw170817_detection}.  No gravitational-wave signal from the merger or post-merger phase was observed~\cite{abbott17_gw170817_postmerger,abbott19_gw170817_postmergerII}.  Approximately \unit[1.7]{s} after the inferred merger time, GRB 170817A was observed by the \textit{Fermi} Gamma-ray Burst Monitor~\cite{GW170817_FermiGBM,FermiGBM} and \textit{Integral}~\cite{GW170817_Integral}, with an optical/UV counterpart detected by an array of instruments less than eleven hours later~\cite{abbott17_gw170817_multimessenger,arcavi17, coulter17,lipunov17,soares-santos17,tanvir17,valenti17}.  Subsequent observations across a majority of the electromagnetic spectrum have continued for more than 1000 days~\citep[e.g.,][]{fong19,hajela19,hajela20,troja20}.

Gravitational-wave observations of the inspiral phase of binary neutron star mergers such as GW170817 and the more-recent GW190425~\cite{abbott20_gw190425} provide valuable insight into the progenitor neutron stars, including their masses and spins, as well as the \textit{cold} equation of state of nuclear matter~\cite{abbott17_gw170817_detection,abbott17_gw170817_gwgrb,abbott19_GW170817_properties,abbott20_gw190425}. The lack of gravitational-wave observations of the merger and post-merger phase limits our inference of the remnant's evolution. We rely instead on indirect observations of the post-merger remnant derived from electromagnetic observations of ejected and stripped material. Understanding the post-merger evolution has the potential to provide valuable, complementary insights into the \textit{hot} nuclear equation of state, as well as details about short gamma-ray bursts and kilonovae hitherto unknown.

This review is dedicated to understanding what we know about binary neutron star post-merger remnants from both an observational and theoretical perspective, and what we hope to learn in the near future as both gravitational-wave and electromagnetic observations increase in both number and detail.

\begin{figure*}[!htbp]
  \begin{tabular}{cc}     
        \includegraphics[width=1.00\textwidth]{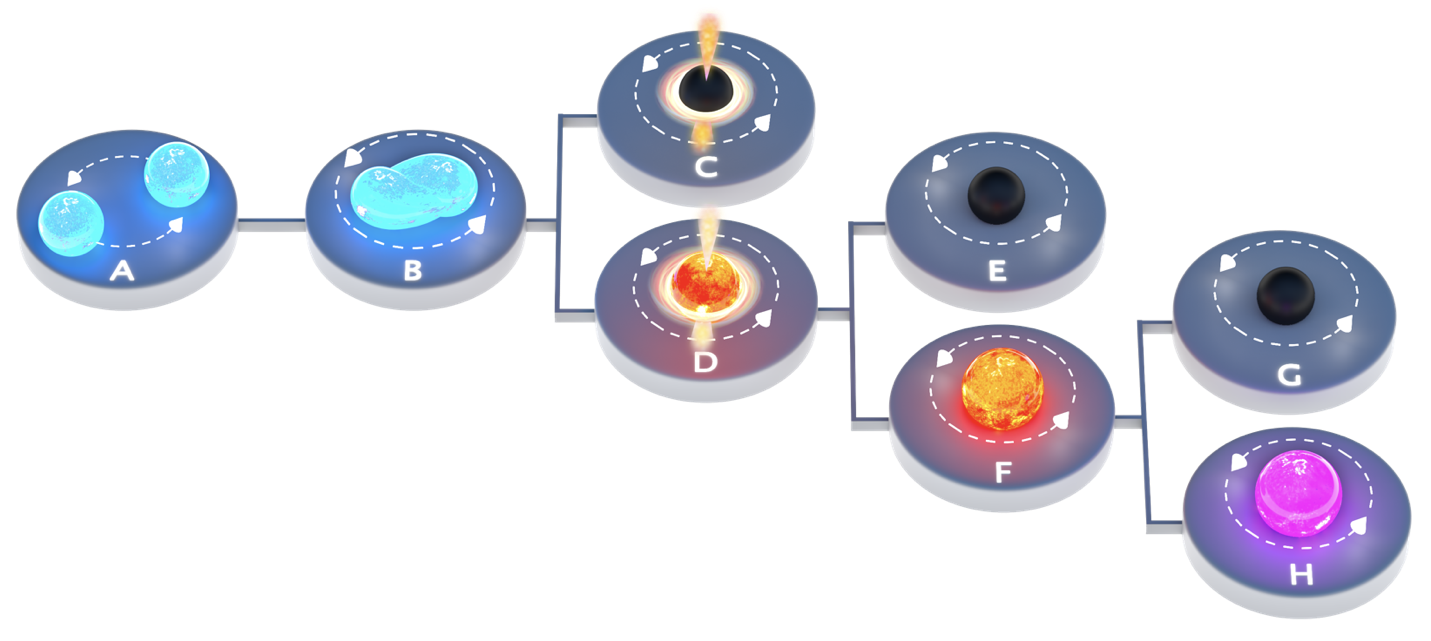}
  \end{tabular}
  \caption{The fate of binary neutron star merger remnants.  Two neutron stars coalesce, losing orbital angular momentum through the emission of gravitational waves until they eventually merge (panels A$\rightarrow$B).  Depending on the mass of the remnant, it will either promptly collapse to form a black hole with an accretion torus and jet (panels B$\rightarrow$C), or form a rapidly differentially-rotating neutron star (panels $B\rightarrow$D).  Depending on the mass of this neutron star, it will either be {\it hypermassive}, in which case it will collapse to form a black hole in $\mathcal{O}(\unit[1]{s})$ (panels D$\rightarrow$E), it will be {\it supramassive}, collapsing to form a black hole in $\lesssim\unit[10^5]{s}$ (panels F$\rightarrow$G), or it will form an infinitely stable neutron star (panels F$\rightarrow$H).}
  \label{fig:post_merger_evolution}
\end{figure*}

There are four possible evolutionary pathways for a neutron star post-merger remnant.  These depend primarily on the remnant mass and the unknown neutron star equation of state.  The latter dictates the Tolman-Oppenheimer-Volkoff mass $M_{\rm TOV}$, which is the maximum mass a non-rotating neutron star can sustain~\cite{tolman39, oppenheimer39}. Observations of pulsars in binary systems indicate $M_{\rm TOV}\gtrsim\unit[2.0]{M_\odot}$~\cite{demorest10,antoniadis13,cromartie19}. Determining the evolutionary pathway of both individual binary neutron star mergers and populations will therefore provide insights into the nuclear equation of state.

Given a remnant mass $M$, the four evolutionary pathways (shown schematically in Fig.~\ref{fig:post_merger_evolution}) are:
\begin{itemize}
    \item $M\gtrsim\chi\,M_{\rm TOV}$: the system promptly collapses to a black hole. Here $\chi$ is the threshold for prompt collapse which is dependent on the equation of state. Most equations of states predict $1.3 \lesssim \chi \lesssim 1.6$~\citep[e.g.,][]{shibata00, shibata06, baiotti17, agathos20, bauswein20} See path A$\rightarrow$B$\rightarrow$C of Fig.~\ref{fig:post_merger_evolution} and Sec.~\ref{sec:prompt_collapse}.
    \item $1.2\,M_{\rm TOV}\lesssim M\lesssim\chi\,M_{\rm TOV}$: a {\it hypermassive} neutron star survives the collision, but collapses to form a black hole on dynamical timescales.  See path A$\rightarrow$B$\rightarrow$D$\rightarrow$E of Fig.~\ref{fig:post_merger_evolution} and Sec.~\ref{sec:short_lived}.
    \item $M_{\rm TOV}<M\lesssim1.2\,M_{\rm TOV}$: a {\it supramassive} neutron star will survive the collision and will collapse to form a black hole on secular timescales.  See path A$\rightarrow$B$\rightarrow$D$\rightarrow$F$\rightarrow$G of Fig.~\ref{fig:post_merger_evolution} and Sec.~\ref{sec:long_lived}.
    \item $M\le M_{\rm TOV}$: a stable neutron star will survive the merger. See path A$\rightarrow$B$\rightarrow$D$\rightarrow$F$\rightarrow$H of Fig.~\ref{fig:post_merger_evolution} and Sec.~\ref{sec:long_lived}
\end{itemize}

Neutron stars can sustain more mass than their non-rotating limit $M_{\rm TOV}$ only when rapidly rotating~\citeg{friedman87, baumgarte00} or extremely hot~\citeg{bauswein10, kaplan14}, in which case either centrifugal support or thermal gradients provide an extra term in the force-balance equation of hydrodynamic equilibrium.  As the star spins down and/or cools, this extra support is lost and the star eventually reaches a point where it can no longer support its own mass and collapses to form a black hole.  In the case of hypermassive stars where $M\gtrsim1.2 M_{\rm TOV}$, uniform rotation cannot provide enough centrifugal support to prevent collapse, implying the star collapses as soon as enough differential rotation is quenched.  While this necessarily happens on the system's dynamical timescale i.e., the free-fall timescale the exact timescale is unknown. We discuss this in detail in Sec.~\ref{sec:short_lived}.

In the supramassive case, even once differential rotation ceases and the star is uniformly rotating, it can still have enough centrifugal support to prevent gravitational collapse.  Secular timescales associated with magnetic dipole radiation and gravitational-wave emission become relevant to establish the timescale for collapse in this case.  It was previously believed that collapse would necessarily happen on a timescale of $\lesssim10^5$ s~\cite{ravi14}, but this is dependent on the strength of the external dipole magnetic field.  Recent afterglow observations of GW170817 may controversially shed new light on this topic~\cite[e.g][]{yu18,li18,piro19,ai20,troja20}.  We discuss this in detail in Sec.~\ref{sec:170817fate}. 

Each of the pathways shown in Fig.~\ref{fig:post_merger_evolution} has different gravitational-wave and electromagnetic signatures, providing hope that one will be able to use such observations to make measurements of the nuclear equation of state.  For example, the specific kilonova color depends on the survival time of merger remnants~\citep[e.g.,][]{li98, metzger10, metzger_fernandez14}.  In this article, we review theoretical and observational aspects of each of the pathways shown in Fig.~\ref{fig:post_merger_evolution}.

The article is set out as follows.  In Sec.~\ref{sec:observational}, we review observational features of GW170817/GRB170817A that potentially allow us to discriminate the post-merger evolutionary pathway.  We detail conflicting reports that independently suggest either a short- or long-lived neutron star survived the merger. There is compelling evidence that we have observed electromagnetic emission from numerous other binary neutron star mergers seen as short gamma-ray bursts.  In Sec.~\ref{sec:otherBNSfate}, we review observational features of short gamma-ray bursts that potentially hint at long-lived neutron star remnants. Following the observational review, we discuss more theoretical aspects of post-merger remnant dynamics, separating the discussion into the different pathways outlined in Fig.~\ref{fig:post_merger_evolution}.  In Sec.~\ref{sec:prompt_collapse} we discuss the prompt formation of black holes, in Sec.~\ref{sec:short_lived} we discuss dynamics and evolution of short-lived hypermassive remnants, and in Sec.~\ref{sec:long_lived} we discuss the evolution of longer-lived supramassive and stable neutron star remnants.

\section{Observational evidence for post-merger remnants}\label{sec:observational}

\subsection{The fate of GW170817}\label{sec:170817fate}
The smoking-gun observation to determine the nature of a post-merger remnant are gravitational waves from the hot, differentially rotating nascent neutron star.
Searches for gravitational waves from possible post-merger remnants of GW170817 or GW190425 have not detected a signal~\citep{abbott17_gw170817_postmerger, abbott19_gw170817_postmergerII,abbott20_gw190425}\footnote{\citet{vanputten19} claim a detection of gravitational waves following GW170817, although see~\citet{oliver19b} for a rebuttal of this work.}. This lack of detection was expected given current detector sensitivities and theoretical models~\citep[e.g.,][and references therein]{clark16,sarin18, zappa18}.  In the following, we concentrate on GW170817 for three primary reasons: first, given GW170817's relative proximity and loudness compared to GW190425; second the fact that the former has a plethora of electromagnetic observations, whereas the latter had no counterpart detections; and third, because of the larger total mass for GW190425, implying it likely promptly collapsed to form a black hole (i.e., path A$\rightarrow$B$\rightarrow$C in Fig.~\ref{fig:post_merger_evolution})~\cite{abbott20_gw190425}.

Although the non-detection of gravitational waves means we are unable to definitively confirm the fate of the post-merger remnant of GW170817, much can be inferred through the various electromagnetic observations, albeit with somewhat conflicting conclusions. Here we elaborate on the possible fates of the post-merger remnant of GW170817\footnote{It is worth noting that the electromagnetic observations of GW170817 are consistent with a neutron star black hole merger, which would produce a black hole remnant. However, to claim GW170817 was a neutron star black hole merger, one must be able to explain the existence of black holes less massive than $2M_{\odot}$.}, the observations that support and contradict each scenario. 

Parameter estimation of the gravitational-wave inspiral signal constrained the total mass of the system to $2.74^{+0.04}_{-0.01} M_\odot$~\cite{abbott19_GW170817_properties}. A small fraction of this total mass~$\approx0.07\,M_\odot$ was ejected and powered the optical kilonova AT2017gfo~\cite{cowperthwaite17, coulter17, soares-santos17, arcavi17, smartt17,chornock17,abbott17_gw170817_multimessenger, troja17_xrays}.  AT2017gfo was first detected as a luminous blue source with a thermal spectrum peaking in optical and ultraviolet frequencies~\citep[e.g.,][]{evans17}, evolving over the course of a few days to become dominated by emission in the near-infrared~\citep[e.g.,][]{tanvir17, pian17}. 
The late-time near-infrared observations agree well with ``red'' kilonova models predicted by the radioactive decay of heavy r-process nuclei~\citep{li98, metzger10}. 
Similarly, the early-time ``blue'' observations are well explained by lower-opacity radioactive material~\citep{metzger10} as would be expected if the outer layers of the ejecta are composed exclusively of light r-process nuclei formed from matter with relatively high electron fractions~\citep{metzger_fernandez14}. We point the interested reader to Refs.~\cite{fernandez16,metzger17} for detailed reviews of kilonovae and Sec.~\ref{sec:short_EM}, where we discuss the theoretical impact of merger remnants on kilonovae.

Different neutron-richness of the material implies different ejecta sources. Two possible sources are the dynamical ejecta launched by tidal forces~\citep{rosswogg99, radice16}, and matter launched from shock heating at the contact boundary of the merger~\citep{bauswein13, hotokezaka13, margalit17}.
The former is ejected along the equatorial plane with typically lower electron fractions than the latter, which is launched along a broad range of directions~\citep[e.g.,][]{sekiguchi16, metzger17}.
Another source of ejecta are outflows from the accretion torus of a central engine~\citep{metzger08, siegel16, siegel18}, typically with broad electron-fraction distributions that can increase with time due to constant neutrino irradiation from a central neutron-star engine. 

As mentioned, observations of AT2017gfo showed evidence for at least two distinctive components: an early-time ``blue'' component and a late-time ``red'' component.
There are also hints at a third ``purple'' component~\cite{villar17}, a point we discuss in more detail below.
Common interpretations of these observations suggest that the two dominant components of the kilonova were powered by two distinct ejecta components. The early blue component by a lanthanide-poor $\approx 0.02 M_{\odot}$ of material with electron fraction $Y_{e} \gtrsim 0.25$ from accretion-disk outflows along the binary polar axis~\citep[e.g.,][]{smartt17, evans17,tanvir17, pian17}, or alternatively from dynamical ejecta launched through shock heating. 
On the other hand, the late-time red component was likely powered by lanthanide-rich $\approx 0.05 M_{\odot}$ ejecta with electron fraction $Y_{e} \lesssim 0.25$~\citep[e.g.,][]{cowperthwaite17}.
The total amount of ejecta and the blue component in particular, offers the first clue into the nature of the post-merger remnant. The amount of ejecta required to produce these observations is incompatible for a remnant that promptly collapsed into a black hole~\citep[e.g.,][]{radice18}. Prompt collapse would have resulted in a primarily red and dimmer kilonova~\citep[e.g.,][]{margalit17, piro19}.

The merger was accompanied by a gamma-ray burst jet that was most-likely structured and off axis~\citep{troja17_xrays, alexander18, mooley18a, mooley18_superluminal, troja19}.
In classical gamma-ray burst models, the jet is launched through accretion onto a black hole, leading to interpretations of the $\unit[1.7]{s}$ delay between the gravitational-wave signal and gamma-ray burst~\citep{GW170817_FermiGRB_paper, GW170817_Integral} as the maximum lifetime of a putative neutron star before it collapses into a black hole~\citep[e.g.,][]{metzger18}. This is further necessitated by claims that the region around the poles needs to be relatively free of ejecta to efficiently launch an ultra-relativistic jet~\citep[e.g.,][]{ciolfi19}, a point we return to in Sec.~\ref{sec:jetlaunch}.
This interpretation is contentious. There are numerous short and long gamma-ray burst observations with evidence of neutron-star central engines~\citep[e.g.,][]{rowlinson10,rowlinson13,lu15,sarin20}, providing observational evidence jets are not only launched from accretion tori around black holes.
In addition, numerical-relativity simulations show short gamma-ray burst jets could potentially be launched given sufficiently large, but realistic, magnetic field strengths of the remnant neutron star ($B \gtrsim\unit[10^{14}]{G}$)~\citep[e.g.,][]{ruiz16_sgrb, ciolfi18_sgrb, mosta20, ruiz21}.  We discuss these points in greater detail in Sec.~\ref{sec:jetlaunch}.
These observations and simulations disfavour the hypothesis that the remnant of GW170817 \textit{must} have collapsed into a black hole in order to launch the jet, opening the possibility for the remnant to be a long-lived supramassive or an infinitely stable neutron star.

With the above caveats in mind, \textit{if} the remnant collapsed to a black hole before launching the jet, strong constraints on the non-rotating neutron star maximum mass can be derived, indicating $M_{\rm TOV}\lesssim\unit[2.2]{M_\odot}$~\cite[e.g.,][]{margalit17,rezzolla18,ai20}. Although other analyses derive a more conservative estimate, $M_{\rm TOV}\lesssim\unit[2.3]{M_\odot}$ by relaxing some of the assumptions on energy emitted in gravitational-wave and neutrinos indirectly imposed by other analyses~\citeg{shibata19, ruiz18_170817}.

While the blue color and total ejecta mass are helpful in ruling out prompt collapse, the exact source of the ejecta mass is unclear. 
Observations suggest an ejecta mass of $\approx 0.02 M_{\odot}$ with a mean velocity and electron fraction of $\approx 0.25c$ and $Y_{e} \gtrsim 0.25$, respectively~\citep{smartt17, evans17,tanvir17, pian17}. 
Although the velocity and high electron fraction agree with a shock-heated dynamical ejecta source, the quantity of material is difficult to explain. 
Relativistic hydrodynamics simulations of equal-mass progenitor systems typically only show ejecta mass $\lesssim 0.01 M_{\odot}$ for soft equations of state that have small neutron star radii $R \lesssim\unit[11]{km}$~\citeg{hotokezaka13, bauswein13, radice16}. 
However, such compact progenitors have less ejecta in the tidal tails, and less mass in the resultant torus, which is inconsistent with the observations of the red component of the kilonova discussed above~\cite{radice18, metzger18}. This raises doubt about the dynamical source of the blue component of the kilonova.
An alternate explanation for the origin of this ejecta posits that it is a magnetized neutrino-irradiated wind from a hypermassive neutron star that survived $\sim\unit[1]{s}$ before collapsing into a black hole and had a strong poloidal magnetic field $B_p \approx \unit[1-3 \times 10^{14}]{G}$~\cite{metzger18}.  Kilonova observations a few hours after the merger, had they existed, could have provided observational support for this hypothesis.

There are other interpretations of the kilonova observations that imply a different fate of the post-merger remnant. 
For example, there is speculation that observations are best fit by a three-component model~\cite{villar17}. Here, the early-time blue kilonova is of similar mass as inferred by other groups, but the late-time observations are dominated by an intermediate-opacity purple component with a significantly weaker red component. 
These purple/red components may be sourced by the accretion disk around a central engine, however it is difficult to interpret why there is a large variation in the opacity of these components, especially if they are coming from the same source.
The purple component may be naturally supported by a long-lived remnant neutron star~\cite{yu18}, where the high-energy emission from the remnant's wind ionizes the surrounding material~\citep{metzger_fernandez14}.
\citet{li18} compared models that could account for both the peak luminosity and time of the kilonova observation, concluding that the observations are best fit with a long-lived neutron star. 

A long-lived remnant supports the low-significance x-ray flare $\unit[155]{d}$ following the merger~\citep{piro19}, potentially due to untwisting toroidal magnetic field similar to x-ray flares from older magnetars~\citep{thompson95,thompson96,piro19} (although, see Ref.~\cite{linDaiGu19} for an alternate explanation). 
Energetic arguments imply the toroidal component of the magnetic field must be $B_{t} \gtrsim\unit[10^{14}]{G}$, albeit with a relatively low $\approx\unit[10^{12}]{G}$ poloidal field such that gravitational-wave emission dominates the spin down for long times~\cite{piro19}.
Such a low poloidal but high toroidal magnetic field structure is perhaps concerning but not dissimilar to the soft gamma-ray repeater SGR 0418+5729~\citep{rea13} (although one should treat such measurements cautiously; see Ref. ~\citep{mastrano13}).
Prolonged x-ray excess almost \unit[1,000]{d} after the merger is still consistent with a long-lived central engine driving the emission~\cite{troja20}.  If indeed this excess is due to a long-lived neutron star, it is almost certainly not supramassive, but likely has a mass below the TOV mass, i.e., it follows path A$\rightarrow$B$\rightarrow$D$\rightarrow$F$\rightarrow$H in Fig.~\ref{fig:post_merger_evolution}. 
Although the above interpretations suggest a long-lived neutron star may have formed in the aftermath of GW170817, there are also problems with this interpretation.
For example, if long-lived, the surface poloidal component of the magnetic field must be $B_p \lesssim\unit[10^{11}-10^{12}]{G}$~\citep{yu18, piro19, ai18}.
This constraint is problematic as it is energetically difficult to launch a Poynting-flux-dominated jet with $B_p \lesssim\unit[10^{14}]{G}$~\citep[e.g.,][and references therein]{ciolfi18_sgrb}. Moreover, Kelvin-Helmholtz and magneto-rotational instabilities dramatically amplify the seed magnetic fields to values greater than $B_{p} \gtrsim \unit[10^{15}]{G}$~\citeg{kiuchi14, kiuchi15, aquileramiret20}.
Furthermore, the lack of signature of the rotational energy of the stable neutron star on the kilonova~\cite{margalit17} and radio~\cite[e.g.,][]{ricci20_radio, schroeder20} implies a considerable amount of energy emitted in gravitational waves, which requires a large ellipticity~\citep{ai20}. If the remnant of GW170817 was long-lived, it perhaps provides the most interesting constraints on the maximum mass of neutron stars $M_{\rm TOV}\gtrsim\unit[2.4]{M_\odot}$~\citep{ai20}.  

If, dear reader, you are not yet confused enough about the nature of the remnant of GW170817, there is one final scenario consistent with all observations.  Namely, the remnant was a supramassive neutron star that spun-down primarily through gravitational waves, collapsing into a black hole after losing centrifugal support approximately \unit[300]{s} after the merger~\citep[e.g.,][]{ai20}. We note that one can impose longer collapse times for different magnetic field configurations.
Such a scenario supports the kilonova observations and the potential lack of observational signature of a rapidly spinning neutron star in other electromagnetic bands at later times. 
In general, one may expect to see the signature of such a remnant on the x-ray afterglow of the short gamma-ray burst~\citep[e.g.,][and Sec.~\ref{sec:otherBNSfate}]{rowlinson10, rowlinson13}.
However, \textit{Swift} did not observe the region until $\sim\unit[0.039]{d}$ after the gamma-ray burst trigger~\cite{evans17}, placing an upper limit on the collapse time $t_{\rm col} \lesssim\unit[0.039]{d}$ of such a supramassive neutron star. If supramassive and collapsing in less than $\sim\unit[0.039]{d}$, the maximum neutron star mass $2.1\lesssim M_{\rm TOV}/M_\odot\lesssim 2.4$~\cite{ai20}.

The first multimessenger binary neutron star merger GW170817 offered an unprecedented opportunity for a detailed study into the the aftermath of such a collision.
Unfortunately, while electromagnetic observations were plentiful and extensive, they remain inconclusive. 
The lack of a smoking-gun observation of gravitational waves from a post-merger remnant make inferring the nature of the remnant difficult. 
In practice, the only scenario everyone seems to agree can be ruled out is of prompt collapse into a black hole. 

\subsection{The fate of other binary neutron star merger remnants}\label{sec:otherBNSfate}
The coincident detection of gravitational waves from a binary neutron star merger GW170817 and the short gamma-ray burst GRB170817A confirmed that at least some of the latter are caused by the former.  
While sensitivity improvements in gravitational-wave detectors will see increased numbers and regularity of binary neutron star merger detections, it will remain true that most observed gravitational-wave signals will \textit{not} be accompanied with electromagnetic signatures.  Likewise, the foreseeable future will see most short gamma-ray burst observations \textit{not} accompanied by gravitational-wave detections.  But there is already a wealth of short gamma-ray burst observational data at our disposal that can, and is, used to understand the remnants of binary neutron star mergers. 

The x-ray afterglows of some short gamma-ray bursts exhibit extended plateaus that indicate energy injection from rapidly rotating neutron star central engines~\cite[e.g.,][]{rowlinson13,lu15}. This even includes the observations of a putative off-axis gamma-ray burst seen only as an x-ray transient CDF-S XT2 that is consistent with these other x-ray afterglows~\cite{xue19, xiao19}.
Such observations are difficult to interpret in the standard afterglow model of synchrotron radiation from shocks produced by a jet interacting with the surrounding interstellar medium. 
The first such observation consistent with energy injection from a short gamma-ray burst was GRB051221A~\cite{fan06}, which was followed by a catalogue of bursts~\cite{rowlinson13}. 
These were shown to be consistent with models of energy injection where the spin down of the nascent neutron star is driven by magnetic dipole radiation~\cite{dai98,zhang01}, a model that has further been extended to include spin down with arbitrary braking indices~\cite{lasky17}, akin to what is seen in the spin down of radio pulsars.

For the majority of short gamma-ray bursts with extended x-ray plateaus, debate continues to rage about the origin of the x-ray flux. For example, evidence for an achromatic jet break in the broadband observations of GRB140903A. This achromatic break has been used to argue that the long-lived emission is due to a combination of jet geometry and dynamics of the fireball~\cite{troja16}, in stark contrast to other works that showed the afterglow is consistent with the spin down of a long-lived central engine~\cite{lasky17, zhang17}.  Systematic Bayesian model comparison between the two scenarios using only x-ray observations overwhelmingly favours the latter explanation~\cite{sarin19}.  

Model comparison between fireball dynamics and a long-lived central engine for another gamma ray-burst GRB130603B yields intriguingly different results~\cite{sarin19}. The discriminator between the two models is a quantity called the odds ratio, but this itself depends on the unknown equation of state. The punch line is that, if the maximum neutron star mass is $M_{\rm TOV}\lesssim\unit[2.3]{M_\odot}$, the data favours the fireball-shock model.  Conversely, if $M_{\rm TOV}\gtrsim\unit[2.3]{M_\odot}$, the data favours the existence of a long-lived neutron star central engine. 
It is worth mentioning the above conclusion relies on knowing the underlying binary neutron star mass distribution, which in light of GW190425, we do not. Moreover, more detailed modeling for each scenario is required, which may further discriminate and potentially yield different conclusions.

The evolution of the x-ray luminosity in these afterglows does allow us to understand somewhat the dynamical evolution of the central engine.  For example, gamma-ray bursts GRB130603B and GRB140903A, are spinning down with braking index $n=2.9\pm0.1$ and $n=2.6\pm0.1$, respectively~\cite{lasky17}, where a braking index of $n=3$ is dipole magnetic spin down in vacuum.  It is worth stressing that all but one radio pulsar with accurately measured braking index falls below $n=3$ where magnetic torques are believed to dominate spindown~\cite{archibald16, clark16b, marshall16}, although see Ref.~\cite{parthasarathy19} for a census of the highly uncertain nature of braking indices in \textit{young} pulsars.  Moreover, it is also worth emphasising that calculations of realistic braking indices for pulsars predicts they fall below 3~\cite[e.g.,][]{melatos97}---we discuss theoretical expectations for the dynamics of neutron-star spin down on these relatively long timescales in Sec.~\ref{sec:long_lived}.

The highly dynamic nature of the newly-born neutron star implies the assumed constant braking index and smooth spin-down evolution are almost certainly naive assumptions. For example, the magnetic field's inclination angle will likely evolve as a function of time, leading to a changing inferred braking index (see~\cite{mus19} for interpretations of long gamma-ray bursts in this context and Sec.~\ref{sec:long_lived} for details of the relevant physics), or the triaxial nature of the remnant may cause precession and a flux-modulated amplitude of the light curve~\cite{melatos00,suvorov20}. In addition, the radiative efficiency is likely not constant throughout the spin down and may, for example, depend on the luminosity of the central engine itself~\cite{xiao19b}. This radiative efficiency likely changes as the shock front decelerates as it ploughs into the interstellar medium~\cite{cohen99,dallosso11,stratta18,sarin20radiative}, and may evolve through plerionic-like emission as electrons fill the cavity within the gamma-ray burst blast wave~\cite{strang19}.  In reality, the dynamical evolution of the remnant and the resultant emission that eventually reaches the observer will be affected by all of these physical processes and more, although which are truly the most dominant is still an open question.

In all, approximately 70\% of short gamma-ray bursts exhibit behaviour consistent with long-lived remnants, split into $\approx30\%$ with supramassive, and $\approx30\%$ stable neutron stars~\cite{gao16}, although these numbers are highly uncertain~\cite[e.g.,][]{margalit19}. In principle, understanding these fractions together with the progenitor mass distribution has the ability to provide strong constraints on the neutron star equation of state through the maximum mass. In practice, our understanding of the numbers are not yet mature enough to make quantifiably-reliable estimates for three reasons. First, determining the nature of the remnant from the x-ray data alone is fraught with difficulties, as highlighted by the various analyses of GRB130603B and GRB140903A mentioned above. Second, while we previously thought we understood the mass distribution of binary neutron stars from galactic radio observations of double neutron star systems~\cite[e.g.,][]{kiziltan13,alsing18,farrow19}, only one of the two extragalactic neutron star mergers has progenitor masses consistent with that distribution~\cite{abbott20_gw190425}. Third, it is possible that some short gamma-ray bursts may be misidentified as long gamma-ray bursts caused by the collapse of massive stars, or that they are caused not by merging neutron stars, but by a neutron star-black hole merger, or that they represent a biased sample of neutron star mergers only including ones that produced a black hole which could launch an ultra-relativistic jet. All of these effects would cause systematic problems with measuring the maximum mass. Understanding both these systematic effects by collecting more gravitational-wave observations to ameliorate the former issue, and more x-ray plateau observations the latter, will eventually provide interesting and stringent constraints on $M_{\rm TOV}$.

Perhaps our best understanding of post-merger behaviour comes from a subset of eighteen short gamma-ray bursts that not only exhibit x-ray plateaus, but also show sudden drops in the x-ray luminosity tens to thousands of seconds after the prompt emission~\cite[e.g.,][]{troja07, rowlinson13, sarin20}.  Such dramatic changes in flux are particularly difficult to explain in the standard fireball-shock scenario, but fit well the premise that a \textit{supramassive} neutron star was born in a neutron star merger and collapses to form a black hole simultaneously shutting of the energy injection.

The supramassive neutron star observations again provide a tantalising way of developing our understanding of the dynamics of the nascent neutron star and the equation of state of nuclear matter~\cite[e.g.,][]{fan13,lasky14,ravi14,li16, gao16,drago16,drago18}. The procedure is straight forward: if we understand the progenitor mass distribution (which we do not), as well as the dominant spin down mechanism (we do not understand that either), and the spin-down rate/braking index (not really), then we can rearrange the set of equations governing the system's evolution to find that the time of collapse is a function of the unknown maximum neutron star mass, which we can therefore infer.  This procedure has been performed a number of times in different works, each arriving at different answers depending on the underlying assumptions at each of the step.  The vanilla assumptions of dipole vacuum spin down of hadronic stars does not well fit the data~\cite{fan13, ravi14}, leading some authors to infer that quark stars, rather than hadronic stars, best explain the data~\cite[e.g.,][]{li16,drago16}, while others infer that gravitational radiation dominates the star's angular momentum loss rather than magnetic dipole radiation~\cite[e.g][]{fan13,gao16}. 

The correct way to do the above procedure rigorously is to include all sources of uncertainty and marginalise over the unknown parameters such as those describing the progenitor mass distribution, the braking index, the equation of state, etc~\cite[for details, see][]{sarin20}. Hierarchical Bayesian inference then allows posterior probability distributions to be calculated for each of these parameters, including those that describe the population as a whole, rather than individual gamma-ray burst afterglows. As one would expect, including all uncertainties implies less-well constrained parameters. Ultimately, the eighteen-known short gamma-ray bursts allow us to constrain $M_{\rm TOV}=\unit[2.31^{+0.36}_{-0.21}]{M_\odot}$, with 68\% uncertainties. Perhaps more interestingly, $69^{+21}_{-39}$\% of remnants are inferred to be spinning down predominantly through gravitational-wave emission, potentially providing interesting consequences for gravitational-wave detection of a post-merger remnant, or indirectly through a stochastic gravitational-wave background. Furthermore, the observations show tentative evidence for these neutron stars to be composed of deconfined quark matter, suggesting a phase transition in the course of merger that may be visible through gravitational-wave measurements of the inspiral~\cite[e.g.,][]{chatziioannou17, bauswein19}.

Having reviewed the observational aspects of short gamma-ray bursts and what can be learned about the remnants of the binary neutron star mergers that power them, we now move onto more theoretical aspects, following the evolutionary scenarios discussed alongside Fig.~\ref{fig:post_merger_evolution}d.

\section{Prompt formation of black holes}\label{sec:prompt_collapse}
\begin{figure}[!htbp]
  \begin{tabular}{cc}     
        \includegraphics[width=0.95\columnwidth]{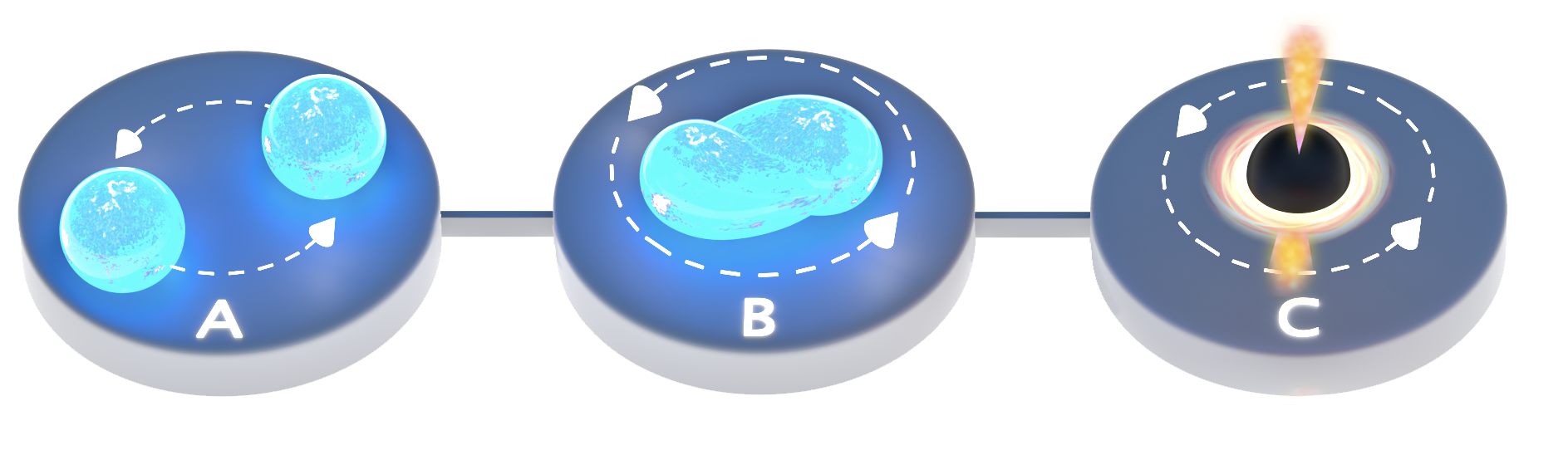}
  \end{tabular}
  \caption{A post-merger remnant of mass $M\gtrsim1.5\,M_{\rm TOV}$ will immediately collapse to form a black hole with an accretion torus and jet.}
  \label{fig:post_merger_evolution_prompt_collapse}
\end{figure}

Perhaps the least interesting of outcomes of a binary neutron star merger is the prompt collapse to a black hole (Fig~\ref{fig:post_merger_evolution_prompt_collapse}; panels B$\rightarrow$C). Upon collapse, a thick accretion torus forms in the black hole's equatorial plane, potentially driving a Blandford-Znajek jet~\cite{blandford77} that is seen as the short gamma-ray burst.  Such a scenario likely leaves an undetectable post-merger gravitational-wave signal, and little direct electromagnetic signal that can be used to infer properties of the central engine, unless the binary has significantly unequal masses, where the tidal disruption of the secondary neutron star can power significant electromagnetic counterparts~\citeg{bernuzzi20_asymmetric}.  

Prompt black hole formation implies that, following coalescence, the gravitational-wave signal simply shuts down, with nearly spherical collapse generating comparatively minimal gravitational-wave emission.  We can approximate the lowest quasi-normal mode ringdown frequency for a remnant black hole as~\cite{echeverria89}
\begin{align}
    f_{\rm gw}^{\rm qnm}\approx\unit[11]{kHz}\left(\frac{M}{3M_\odot}\right)^{-1}\left[1-0.63\left(1-a\right)^{3/10}\right],
\end{align}
where $M$ is the remnant mass and $a$ the dimensionless spin.  Faster rotating black holes emit higher frequency gravitational-wave signals; a non-rotating ($a=0$) black hole of $M\approx3\,M_\odot$ emits its lowest quasinormal mode signal at $f_{\rm gw}^{\rm qnm}\approx\unit[4]{kHz}\,$.  Prompt collapse will more typically result in a remnant with $a\sim0.7$--$0.8$ for which $f_{\rm gw}^{\rm qnm}\gtrsim\unit[6]{kHz}$.  In these regimes, the sensitivity of current interferometers~\cite{abbott16_LRR,LIGO,Virgo} and even proposed future detectors~\cite[e.g.,][]{Aplus, EinsteinTelescope, CosmicExplorer, martynov19, OzHF20} is not sufficient to detect such a signal at relevant distances.

Potentially, the prompt formation of a black hole can have implications for electromagnetic emission, in particular through a lack of sustained energy injection into the x-ray and optical afterglow signal and in the amount of ejecta both dynamical and from an accretion disk.  For example, kilonova observations would likely differ from those of GW170817/AT2017gfo in that they would be primarily red due to the lack of neutrino irradiation of the ejected material and inferred to have less mass~\cite{margalit17, piro19}.  If observations can concretely say no remnant survived the collisions, tight constraints could be placed on the equation of state of nuclear matter.  We return to this in subsequent sections. 

\section{Short-lived, hypermassive neutron stars}\label{sec:short_lived}

\begin{figure}[!htbp]
  \begin{tabular}{cc}     
        \includegraphics[width=0.95\columnwidth]{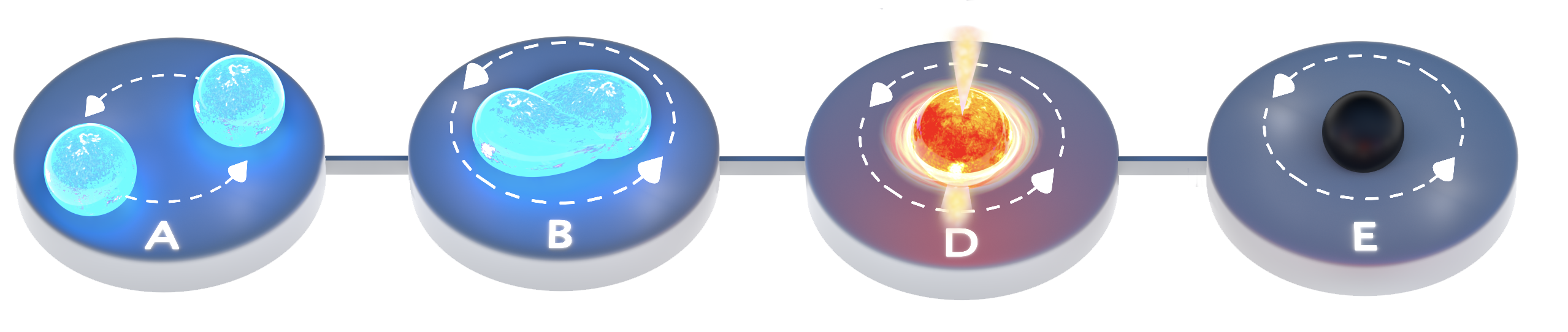}
  \end{tabular}
  \caption{A post-merger remnant of mass $1.2\,M_{\rm TOV}\gtrsim M\gtrsim1.5\,M_{\rm TOV}$ will form a hypermassive neutron star which will collapse to a black hole on a timescale~$\mathcal{O}(\unit[1]{s})$.}
  \label{fig:post_merger_evolution_hypermassive}
\end{figure}

A dominant fraction of binary neutron star mergers likely form post-merger neutron star remnants that are either hypermassive, supramassive, or long-lived~\cite[e.g.,][]{gao16, margalit19}.  In all these cases, the remnant will undergo a short period of highly-dynamic activity (Fig.~\ref{fig:post_merger_evolution_hypermassive}; panels B$\rightarrow$D) before either settling down into rigid-body rotation (Fig.~\ref{fig:post_merger_evolution}; panels D$\rightarrow$F), or collapsing to form a black hole (Fig.~\ref{fig:post_merger_evolution_hypermassive}; panels D$\rightarrow$E).  The latter scenario describes that of a hypermassive neutron star.  We review those dynamics in Sec.~\ref{sec:short_dynamics}, and consequences for electromagnetic and gravitational-wave observations in Secs.~\ref{sec:short_EM} and~\ref{sec:short_GW}, respectively.  

We note that the dynamics and emission scenarios outlined in this section are also relevant for the early evolution of a supramassive (Fig. 1: panels D$\rightarrow$F$\rightarrow$G) and an eternally-stable neutron star (panels D$\rightarrow$F$\rightarrow$H).

\subsection{Dynamics and the Collapse Time}\label{sec:short_dynamics}
The short-term dynamics of binary neutron star merger remnants in the first tens of milliseconds and up to a few seconds depends on physics as yet not completely understood.
Immediately following the merger, the rotational profile of the remnant and the amount of mass in the disk depend principally on the mass ratio of the progenitor~\citeg{oechslin06, fernandez13, metzger_fernandez14, bernuzzi20} and the equation of state~\citeg{shibata06_diskmass, shibata06, kastaun15}.  The hot remnant is rotating differentially, has an incredibly strong magnetic field, several large-amplitude oscillation modes, and cools primarily through neutrino emission.  If the mass of the remnant is above the supramassive limit, then it will collapse within a few seconds of formation.  General relativistic magnetohydrodynamic simulations are currently inadequate to robustly and reliably determine that lifetime for several reasons.  First, not all of the aforementioned physics is adequately resolved~\cite[][and see below]{kiuchi18}, and second, the simulations generally only last $\lesssim\unit[50]{ms}$~\cite{ciolfi17} (although~\cite[see e.g.,][]{ciolfi19, ciolfi20_250ms, shibata21} for recent, long-lived simulations lasting up to \unit[1]{s} post merger).  
 
Differential rotation of the remnant and thermal pressure are expected to provide extra centrifugal support to sustain the remnant above the maximum rigidly-rotating mass limit.  When either of these is sufficiently quenched, the remnant will collapse rapidly to form a black hole.  Conventional wisdom~\cite[e.g.,][]{baumgarte00,shapiro00} states that differential rotation is suppressed on an Alfv\'en timescale, which can be approximately written as
\begin{align}
    \tau_{\rm A}\approx\unit[0.3]{s}\left(\frac{\left<B\right>}{\unit[10^{15}]{G}}\right)\left(\frac{M}{2M_\odot}\right)^{1/2}\left(\frac{R}{\unit[10]{km}}\right)^{-1/2},\label{eq:Alfven}
\end{align}
where $\left<B\right>$ is the volume-averaged magnetic field inside the star, and $R$ is the star's radius.  Below we argue that magnetic-field quenching should occur on a longer timescale than expressed in Eq.~\ref{eq:Alfven}. Clearly, not all of the differential rotation must be suppressed for the remnant to collapse, but only the critical threshold must be reached which will depend ultimately on a number of factors, including how much more massive the remnant is than the supramassive-mass threshold.

The speed with which the internal magnetic field grows, its saturation strength, and the role of magnetic-field instabilities are potentially the most significant unknowns when considering the suppression of differential rotation~\citep[e.g.,][]{ferrario15}.  General relativistic, three-dimensional magnetohydrodynamic simulations of mergers show the magnetic field at the shock interface between the two stars grows rapidly at initial times~\cite[e.g.,][]{giacomazzo13,kiuchi14,kiuchi15,ruiz16_sgrb, ciolfi17,kiuchi18,ciolfi19, aquileramiret20}. 
This field growth is primarily due to the Kelvin-Helmholtz instability that develops at the shear boundary between the two stars. The highest-resolution simulations show amplification of the average field up to $\sim10^3$ times the seed field~\cite{kiuchi18}, although this is still under-resolved and considered a lower limit on the potential of the Kelvin-Helmholtz instability~\cite{kiuchi18,duez19}. Recently, numerical simulations with a resolution $\sim\unit[37]{m}$ (cf. typical resolutions $\gtrsim\unit[120]{m}$) show amplifications up to $10^{5}$ times the seed field~\citep{aquileramiret20}. This resolution is considered adequate for resolving the Kelvin-Helmholtz instability.
Magnetic winding, the magneto-rotational instability, and turbulence can subsequently amplify the field to~$\gtrsim\unit[10^{16}]{G}$ on short timescales~\citeg{duez06a, duez06b, zrake13, siegel13}.  It is worth noting that these latter effects are beyond the resolution limit of current numerical simulations~\cite{baiotti17,kiuchi18}, in particular with respect to MRI-driven turbulence, implying quantitative evolution of the star's magnetic-field growth should not be trusted.  

Most numerical-relativity simulations are performed under the assumption of ideal magnetohydrodynamics, where infinite conductivity of the fluid implies the magnetic field is frozen into the dynamic fluid. However, the hot remnant should almost certainly have regions of finite conductivity, implying flux freezing is not the correct assumption. The scarce numerical-relativity simulations of post-merger remnants that \textit{include} the effects of resistive magnetohydrodynamics in the core indeed show the expected result that the magnetic field lags behind the fluid~\cite{dionysopoulou15}.  This implies that magnetic winding is less efficient than previously believed, and the Alfv\'en timescale of Eq.~\ref{eq:Alfven} should be taken as a rough \textit{lower} bound for the quenching of differential rotation through magnetic-field winding. 

In addition to differential rotation, the nascent star is supported through thermal pressures.  Merger simulations ubiquitously show temperatures at the shock interface in excess of $\unit[3\times 10^{11}]{K}$, and the majority of the star above $\unit[10^{11}]{K}$~\cite[e.g.,][]{sekiguchi11a,sekiguchi11b,foucart16, perego19}.  Simulations that take cooling into account show the remnant may not collapse for a few cooling timescales~\cite{paschalidis12}, which may be as long as a few seconds~\cite{duez19}.  We return to the cooling of the nascent neutron star in Sec.~\ref{sec:thermal_evolution}, but the above argument suggests the dominant physical processes dictating the collapse time of the remnant is magnetic winding rather than thermal dissipation.

Quite clearly there is complicated physics dictating the collapse time of a hypermassive remnant.  If one believes that the collapse of the hypermassive star is required to launch the gamma-ray burst, then \unit[1.74]{s} delay between the gravitational-wave inferred merger time of GW170817 and the gamma-ray detection could be partially explained by the collapse time of the hypermassive star. This is complicated by the additional unknown time it takes for the jet to launch from the black hole, and for the jet to break out from the merger ejecta.  It would therefore be nice to be able to directly measure the collapse time of the hypermassive star, which may be done with future gravitational-wave detections---see Sec.~\ref{sec:short_GW}.

Ultimately, when the remnant collapses to form a black hole, it will do so approximately on a dynamical timescale~\cite{stark85}. The dynamics and timescale of the collapse depend heavily on the system's angular momentum~\cite{baiotti05,baiotti07} and the degree of differential rotation~\cite{giacomazzo11}.  Although the free-fall timescale of a massive neutron star is $\lesssim\unit[0.01]{ms}$, the collapse timescale for a rapidly, differentially rotating neutron star is $\sim\unit[1]{ms}$~\cite{giacomazzo11}.  Whether the collapse time can be measured upon a successful detection of post-merger gravitational waves is an open question (see Sec.~\ref{sec:short_GW}), however the lifetime of the remnant is believed to indirectly impact the spectral evolution of the electromagnetic signature; a point to which we now turn.

\subsection{Electromagnetic consequences}\label{sec:short_EM}
Kilonova emission is a direct result of radioactive decay of heavy elements produced by the merger ejecta.  
As this review deals specifically with the merger remnant, rather than the merger itself, we do not review the physics of kilonovae emission directly. 
Instead, we focus on the effects the hypermassive remnant's lifetime, dynamics, and evolution have on the kilonova and other electromagnetic emission channels.  For reviews of kilonova itself, see e.g., Refs.~\cite{fernandez16, metzger17}.

The ejecta in a neutron star merger can be broadly split into two categories: dynamical ejecta produced in the merger itself, and the (secular) outflow from the accretion disk formed around the remnant object. We note that there may be additional sub-dominant channels that contribute to the total ejecta.
Both the amount of ejecta and its properties (velocity and electron fraction) are intrinsically connected to binary parameters and the fate of the merger remnant, with the electron fraction being perhaps the most critical as it directly impacts what elements can be synthesised and therefore the color of the kilonova.  

Dynamical ejecta usually constitutes two sources: shock-heated ejecta from the contact interface between the two merging neutron stars and spiral arms from the tidal interactions in the merger. The former following a $\sin^{2} \theta$ distribution with respect to the polar angle~\citeg{perego17}, while the latter is launched predominantly in the equatorial plane~\citeg{bernuzzi20}.
The total quantity of dynamical ejecta depends sensitively on the fate of the merger remnant and the binary mass ratio~\citeg{bauswein13b, lehner16}, if the remnant promptly collapses into a black hole, there will be little shock-heated ejecta, as the region is promptly swallowed up~\cite{bauswein13b, ciolfi17, radice18b}. In general, asymmetric binaries tend to produce more ejecta~\cite{rezzola10,bauswein13b}. However, we emphasise that this relationship is not well understood quantitatively. Numerical simulations suggest that the total dynamical ejecta in a merger is in the range of $10^{-4}-10^{-2} M_{\odot}$ with velocities in the range $0.1-0.3$c~\cite[see e.g.,][]{hotokezaka13} and a broad electron fraction distribution, $Y_{e} \sim 0.1 - 0.4$~\cite{radice16} which dictates what elements can be synthesised from this ejecta, potentially up to an atomic mass number, $A\sim195$~\cite{wanajo14,martin15}. We note that this is an area of active research, with significant uncertainties in many critical nuclear reaction quantities~\cite{zhu20}.

The other source of ejecta in a neutron star merger is the outflow from the accretion disk that forms around the remnant object. The quantity of mass in the accretion disk range from $\sim 0.01-0.3 M_{\odot}$ depending on the binary parameters~\citep[e.g.,][]{oechslin06} and the fate of the merger remnant, with prompt formation likely resulting in less mass around the remnant object~\cite[e.g.,][]{perego14, metzger_fernandez14,martin15, metzger17}. Depending on the lifetime of the remnant neutron star (as we discuss below), outflows from this disk likely contribute more mass to the ejecta than the dynamical ejecta launched in the merger itself~\citep{perego14, fernandez16, siegel18}, a statement seemingly verified by the inferred properties of the kilonova, AT2017gfo~\cite[e.g.,][]{smartt17,kasen17, metzger18}. It is the properties of this outflow that are most affected by the nature and lifetime of the remnant and make the biggest impact on the kilonova. In particular, the cooling of the nascent neutron star through neutrino losses. We discuss the thermal evolution of neutron stars in detail in Sec~\ref{sec:thermal_evolution}.

The accretion disk itself evolves viscously and cools through neutrino emission, driving a wind similar to proto-neutron stars born in supernovae~\cite[e.g.,][]{beloborodov08, metzger08}. The mass loss through this channel is dependent on the neutrino flux, which as we elaborate below is connected to the fate of the remnant. Depending on the merger outcome, this process results in a mass loss of up to $10^{-3} M_{\odot}$ either from the disk, remnant neutron star or a combination of the two. Further evolution of the disk is dictated by angular momentum transport, either through turbulence generated by the magneto-rotational instability~\citep[e.g.,][]{metzger_fernandez14, siegel17, siegel18} or by spiral density waves excited by oscillations of a neutron star remnant which expand the disk outwards~\citep[e.g.,][]{nedora19, metzger17}. Initially, the disk accretes matter at a relatively high rate, but once this rate drops below a critical threshold, cooling through neutrinos is ineffective and the disk thermally expands. In this process, free nucleons recombine into $\alpha-$particles which releases enough energy to unbind a significant fraction of the disk~\citep[e.g.,][]{beloborodov08, metzger08, metzger10, fernandez13, perego14, martin15, metzger_fernandez14, siegel17, fernandes19}. The amount of ejecta this process unbinds is again connected to the fate of the remnant, with numerical simulations suggesting prompt black hole formation unbinds up to $\sim 40\%$ of the disk~\cite[e.g.,][]{fernandez16} while a neutron star remnant unbinds potentially up to $\sim 90\%$~\cite[e.g.,][]{siegel17} due in large part to the additional neutrino flux from the remnant neutron star. 

In principle, the quantity of ejecta driven by the outflow of the accretion disk is closely linked to the fate of the remnant and can principle be used to infer the fate of the remnant. However, there are substantial quantitative uncertainties~\citep[e.g.,][]{bernuzzi20} due to simplified neutrino treatments, numerical artifacts from limited resolution, and additional unmodelled processes such as a magnetised neutrino driven wind~\citep[e.g.,][]{metzger18}. Moreover, there are significant systematic uncertainties associated with nuclear reaction networks, opacities, etc., that can led to substantial bias in inferring properties of the kilonova from observations~\cite{zhu20, barnes20}. 

Ignoring the aforementioned complications, a more reliable discriminator of the fate of the remnant is to consider the impact of neutrino radiation on the electron fraction of the ejecta. As we discuss in more detail in Sec~\ref{sec:thermal_evolution}, nascent neutron stars cool by emitting neutrinos. The additional neutrino flux increases the electron fraction with time through $\nu_{e} + n \rightarrow p + e^{-}$. The electron fraction continues to increase with longer remnant lifetimes making it increasingly difficult to synthesize heavier $r$-process elements~\cite[e.g.,][]{metzger_fernandez14, kasen15, lippuner17, kyohei20_diversity}, which directly affects the colour of the kilonova. For $Y_{e} \lesssim 0.25$, a predominantly red kilonova is expected with elements greater than atomic mass $A \gtrsim 140$, while for electron fractions $Y_{e} \gtrsim 0.25$, lighter elements are expected and the colour of the kilonova is predominantly blue~\citep[e.g.,][]{metzger17}. Numerical calculations suggest that a remnant lifetime longer than $\sim 300$ ms will make $Y_{e} \gtrsim 0.25$ in the ejecta from the disk outflow~\cite{lippuner17}, although other calculations suggest a larger lifetime of $\sim 1$s for a similar electron fraction~\cite{sekiguchi16, kyohei20_diversity}.

\subsection{Gravitational-wave emission and detection}\label{sec:short_GW}
Gravitational-wave emission from the first second post-merger is expected to have a relatively large strain amplitude, possibly comparable to the peak amplitude of the inspiral phase, albeit with frequencies in the kHz range. This makes them an interesting target for current and future ground-based gravitational-wave observatories.  In the following two subsections we review state-of-the-art predictions for gravitational-wave emission and detection, respectively.

\subsubsection{Gravitational-wave emission}
Numerical-relativity simulations of binary neutron star mergers and their subsequent post-merger evolution show that gravitational-wave emission from a hypermassive remnant is dominated by the quadrupolar $f$-mode~\cite[e.g.,][]{xing94,ruffert96}.  Depending on the equation of state, this occurs anywhere from $\sim2$ to $\unit[4]{kHz}$~\cite[e.g.,][]{takami15}, and is strongly correlated with the star's compactness and tidal deformability~\cite[e.g.,][]{bauswein12, bauswein_janka12, hotokezaka13,read13,takami14,bauswein19b}.  It is actually somewhat surprising that the frequency of the dominant $f$-mode post-merger correlates so well with the tidal deformability and compactness---these are quantities calculated for \textit{cold, non-rotating} neutron stars, whereas the hypermassive post-merger remnant is rapidly rotating and has a temperature~$\gtrsim\unit[{\rm few}\times10^{10}]{K}$~\cite[e.g.,][]{sekiguchi11a,sekiguchi11b,foucart16}.  This suggests rotational and temperature effects play a minor role in the dominant properties of the gravitational-wave signal, and broadly implies that a successful measurement of the dominant post-merger gravitational-wave frequency is a robust measurement of the nuclear equation of state.  Importantly, if the gravitational-wave frequency of the post-merger oscillations do \textit{not} match up to the tidal deformation measured from the inspiral phase, this could be the signature of a first-order hadron-quark phase transition occurring in the core of neutron stars at high temperatures and pressures~\citeg{most19,bauswein19}.

\begin{figure}[!htbp]
  \begin{tabular}{cc}     
        \includegraphics[width=1.0\columnwidth]{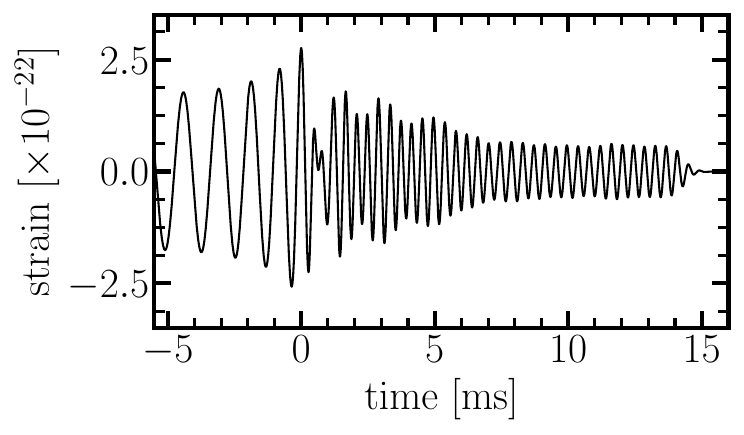}
  \end{tabular}
  \caption{Gravitational-wave strain from a numerical relativity simulation of a binary neutron star merger.  Gravitational-waves from the hypermassive post-merger remnant can have amplitudes comparable to that of the peak of the inspiral.  In this example, the gravitational-wave signal shuts off rapidly approximately $\unit[14]{ms}$ after the merger, signifying the collapse of the hypermassive neutron star to a black hole.  This simulation~\cite{bernuzzi14,dietrich18} is of an equal-mass binary with $M=\unit[1.35]{M_\odot}$ component masses using the H4 equation of state at a distance of $\unit[40]{Mpc}$.}
  \label{fig:HMNSWaveform}
\end{figure}

In Fig.~\ref{fig:HMNSWaveform} we plot an example gravitational waveform from a hypermassive post-merger remnant~\cite{bernuzzi14}\footnote{This waveform is publicly available through the CoRe database of binary neutron star merger waveforms~\cite[waveform ID BAM:0035;][]{dietrich18}}. This shows the merger of two $\unit[1.35]{M_\odot}$ neutron stars with the H4 equation of state~\cite{lackey06} at a distance of $\unit[40]{Mpc}$.  The strain amplitude is maximal at time $t=0$, which we take as a proxy for the merger time (other definitions are often used, including the first contact between the two stars).  For this simulation, the neutron star collapsed after time $t\approx\unit[14]{ms}$, although we reiterate this collapse time is not a reliable prediction for all of the reasons given in Sec.~\ref{sec:short_dynamics}.

\begin{figure}[!htbp]
  \begin{tabular}{cc}     
        \includegraphics[width=1.0\columnwidth]{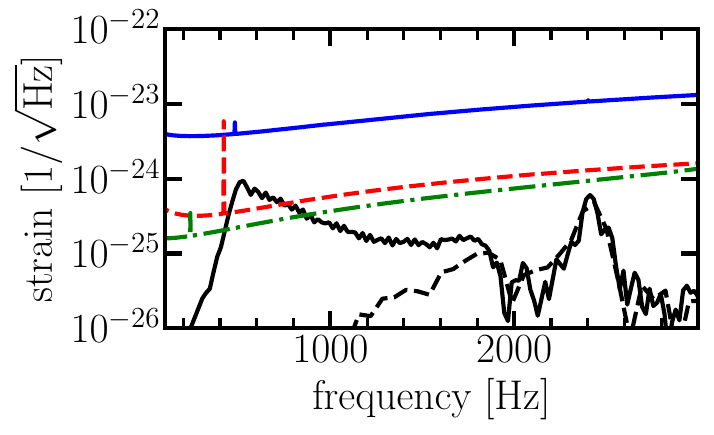}
  \end{tabular}
  \caption{Gravitational-wave amplitude spectrum from the hypermassive post-merger remnant shown in Fig.~\ref{fig:HMNSWaveform}.  The solid black curve shows the full spectral density including almost $\unit[30]{ms}$ if inspiral, whereas the dashed black curve shows the spectral density of just the post-merger remnant (i.e., truncating the time series at the merger).  The solid blue, dashed red, and dot-dashed green curves are the projected amplitude spectral densities of Advanced LIGO at design sensitivity, the Einstein Telescope, and Cosmic Explorer, respectively.  This post-merger signal has a single-detector signal-to-noise ratio of 1.0, 9.0 and 12.7 for the three detectors, respectively.  }
  \label{fig:HMNSWaveformFT}
\end{figure}

In Fig.~\ref{fig:HMNSWaveformFT} we plot the gravitational-wave amplitude spectral density for the waveform shown in Fig.~\ref{fig:HMNSWaveform}.  The amplitude spectra of the full waveform, which includes almost ten full orbits of the binary prior to merger, is shown as the solid black curve, whereas the dashed black curve includes only the post-merger component.  For comparison, we also plot the design amplitude noise spectral densities for three instruments; Advanced LIGO~\cite[solid blue curve;][]{LIGO}, the Einstein Telescope~\cite[red dashed curve;][]{EinsteinTelescope} and Cosmic Explorer~\cite[dot-dashed green curve;][]{CosmicExplorer}.  

For the three sensitivity curves shown alongside the predicted gravitational-wave spectrum in Fig.~\ref{fig:HMNSWaveformFT} we can calculate the expectation value of the single-detector, optimal matched filter
\begin{align}
    \left<{\rm S/N}\right>^2=4{\rm Re}\int df\frac{\left|\tilde{h}(f)^2\right|}{S_{h}(f)},
\end{align}
where $\tilde{h}(f)$ is the Fourier transform of the gravitational-wave time series, and $S_h(f)$ is the noise power spectral density.  We calculate this for the post-merger signal only (i.e., for time $t>0$) and find $\left<{\rm S/N}\right>=1.0$, 9.0, and 12.7, for Advanced LIGO, Einstein Telescope, and Cosmic Explorer, respectively. 

Together with the latest prediction for the merger rates derived from the first gravitational-wave observation of a binary neutron star merger~\cite{abbott17_gw170817_detection}, one can estimate the expected event rates for post-merger gravitational-wave detection.  Marginalising over a range of equations of state, Ref.~\cite{martynov19} predicted~$\lesssim2$, ${\rm S/N}>5$ detections of a post-merger remnant per year with Einstein Telescope, and~$\lesssim10$ such detections per year with Cosmic Explorer~\cite[see also][]{clark16}.  According to these calculations, the pay-off is likely to only come with third-generation detectors where the high-frequency ($\gtrsim$kHz) sensitivity is increased by a factor of at least ten over Advanced LIGO design sensitivity.  This has prompted many to also think about the potential for stacking multiple sub-threshold post-merger signals with second- or third-generation inteferometers~\cite{yang18}, or to build dedicated high-frequency gravitational-wave instruments with the primary science goal to detect tidal effects during the inspiral of binary neutron star mergers as well as their post-merger remnants~\cite[e.g.,][]{miao18,martynov19,OzHF20}.

When the hypermassive star collapses to form a black hole, it does so on a relatively short timescale ($\sim\unit{ms}$; see Sec.~\ref{sec:short_dynamics} and Fig.~\ref{fig:HMNSWaveform}), implying the signal is potentially upward of $\sim\unit[1]{kHz}$.  However, the amplitude of the signal is weak; optimistic estimates from simulations of collapsing, differentially-rotating stars suggest they may be detectable at a distance of up to $\unit[10]{Mpc}$ with third-generation detectors such as the Einstein Telescope or Cosmic Explorer~\cite{giacomazzo11}.  Numerical relativity simulations of binary neutron star mergers with subsequent hypermassive star formation and collapse seem to show no discernible burst of radiation above that seen from the star's oscillations immediately prior to collapse.

It is currently an open question as to whether gravitational-wave parameter estimation methods targeting hypermassive neutron star signals---see next section---can infer the collapse time with a successful  gravitational-wave detection.  If they can, this would allow us to probe the complex physics that governs the quenching of differential rotation discussed in Sec.~\ref{sec:short_dynamics}.

\subsubsection{Gravitational-wave detection methods}
The LIGO/Virgo collaborations have a number of methods in place to search for and characterise gravitational waves in the immediate aftermath of a neutron-star merger.  Two algorithms were used to search for short-lived gravitational-wave signals following GW170817~\cite[]{abbott17_gw170817_postmerger, abbott19_GW170817_properties}: Coherent Wave Burst~\cite[cWB;][]{klimenko16} and BayesWave~\cite{cornish15,littenberg15,chatziioannou17}.  The cWB algorithm searches for coherent excess power in multi-resolution wavelet transformations, while BayesWave uses Bayesian inference, modelling the gravitational-wave signal itself as a linear superposition of wavelets; see Refs.~\cite{abbott17_gw170817_postmerger} and~\cite{abbott19_GW170817_properties}, respectively, for details of the specific implementation and setup of the two algorithms searching for gravitational waves following GW170817.  Neither methods found any hint of a signal, but placed upper limits on the total energy emitted in gravitational waves.

The two methods described above do not rely on waveform models to search for the gravitational-wave signal. While they are therefore more robust than modelled searches that use template waveforms, they are also less sensitive~\cite[][]{tsang19,easter20}.  However, modelled searches are in their infancy due to a paucity of enough reliable gravitational-wave templates to perform matched-filter searches.  This is rapidly changing, with analytic approximations~\cite{bauswein16,bose18}, principal component decompositions~\cite{clark16}, and machine-learning algorithms~\cite{easter19} showing promising results fitting to numerical-relativity waveforms.  

Two recent Bayesian methods have been independently developed that use analytic waveforms.  Using only a single oscillation mode modelled as a damped sinusoid (or Lorentzian function in the frequency domain), \citet{tsang19} showed an average mismatch between numerical-relativity injections and recovered signals of 0.15.  A single damped sinusoid allows the main $f_2$ peak to be measured, which is the key peak for determining the equation of state.  This method is therefore capable of distinguishing inspiral and post-merger inferences of the equation of state~\cite{tsang19} to, for example, determine potential quark deconfinement in the stellar core that only occurs at high temperatures~\cite{bauswein19}.  

Instead of a single damped sinusoid,~\citet{easter20} modelled the full waveform as a linear sum of three damped sinusoids~\cite[inspired by Refs.][]{bauswein16,bose18}, also allowing all three frequencies to drift linearly in time.  They found an average mismatch of only 0.03, implying the method is $\approx15\%$ more sensitive than Ref.~\cite{tsang19}.  The addition of the extra mode oscillations in the analytic waveform approximations is unlikely to improve equation of state estimates, however realistic modelling of the frequency drift of the fundamental $f_2$ mode may have some, as yet undetermined effects.  

In reality, given the potential inaccuracies of gravitational waveforms from numerical-relativity simulations (see previous section), both modelled and unmodelled searches will be needed as this field hopefully moves from the development to the observational stage.

Of course, many binary neutron star coalescences will likely be detected before a bona fide post-merger detection.  This opens the possibility to effectively \textit{stack} sub-threshold events by either multiplying Bayes factors from individual events or by coherent summation of signals relying on pre-merger phase information~\cite{yang18}.

\section{Long-lived neutron star remnants}\label{sec:long_lived}
\begin{figure}[!htbp]
  \begin{tabular}{cc}     
        \includegraphics[width=0.95\columnwidth]{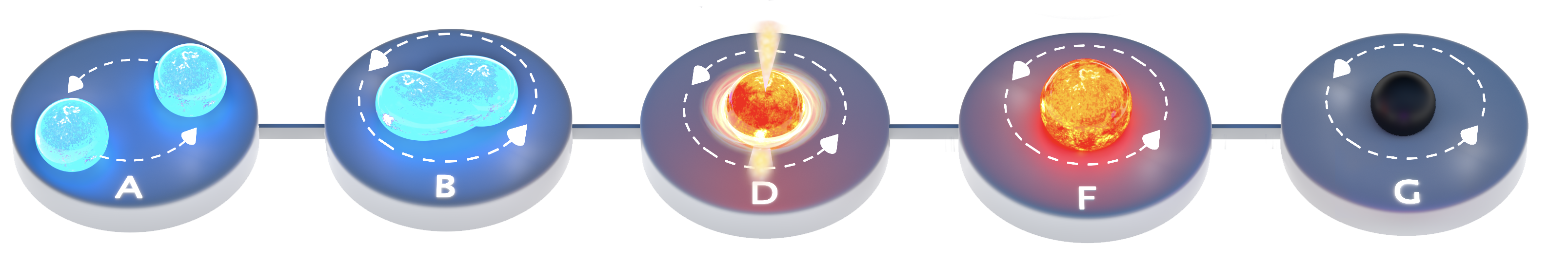}
  \end{tabular}
  \caption{A post-merger remnant of mass $1.0\,M_{\rm TOV}\le M\gtrsim1.2\,M_{\rm TOV}$ will form a supramassive neutron star, which will collapse to a black hole on a timescale~$\lesssim\unit[10^5]{s}$}
  \label{fig:post_merger_evolution_supramassive}
\end{figure}

\begin{figure}[!htbp]
  \begin{tabular}{cc}     
        \includegraphics[width=0.95\columnwidth]{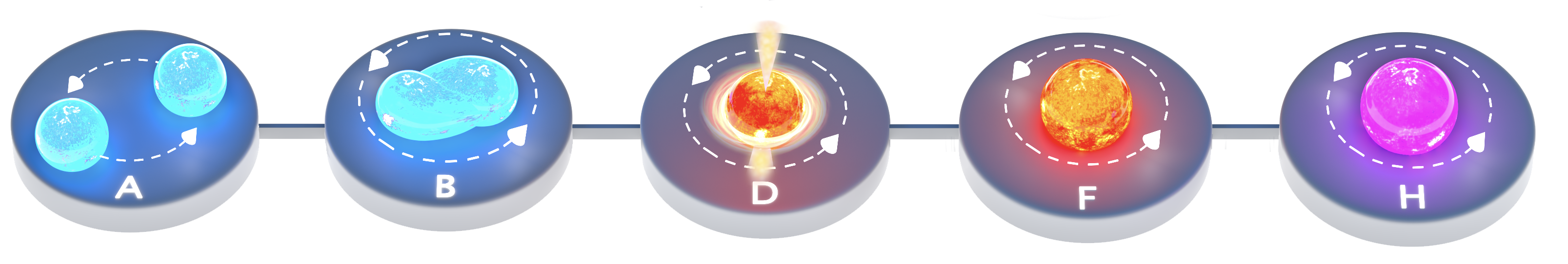}
  \end{tabular}
  \caption{A post-merger remnant of mass $M\le 1\,M_{\rm TOV}$ will form an infinitely-stable neutron star.}
  \label{fig:post_merger_evolution_stable}
\end{figure}

Post-merger remnants less massive than $\sim1.2M_{\rm TOV}$ will survive for more than one second.  They are deemed supramassive if their mass is greater than the non-rotating limit $M_{\rm TOV}$---Fig.~\ref{fig:post_merger_evolution_supramassive}---or infinitely stable if their mass is less than $M_{\rm TOV}$---Fig.~\ref{fig:post_merger_evolution_stable}.  Observational evidence from the x-ray afterglow of short gamma-ray bursts suggests a non-negligible fraction of binary neutron star mergers may result in these outcomes~\cite[e.g.,][]{rowlinson13,lu15}.  

Figures~\ref{fig:post_merger_evolution_supramassive} and \ref{fig:post_merger_evolution_stable} emphasise an important point: even supramassive and stable remnants will still undergo a period of strong differential rotation immediately after the merger (panel B), implying the gravitational-wave emission and detection discussion from Secs.~\ref{sec:short_dynamics} and \ref{sec:short_GW} are still relevant.  However, electromagnetic emission due to r-process nucleosynthesis discussed in Sec.~\ref{sec:short_EM} may be different depending on the lifetime of the remnant; a point we return to in Sec.~\ref{sec:long_obs}.

\subsection{Can a long-lived neutron star launch a jet?}\label{sec:jetlaunch}
The question whether a neutron star can launch an ultra-relativistic jet that can produce a short gamma-ray burst continues to engage theorists and observers. Before we review our more recent understanding of this question, it is intriguing to start near the beginning and briefly review our understanding of short gamma-ray bursts themselves. We refer the interested reader to more detailed reviews of gamma-ray bursts~\cite[e.g.,][]{nakar07,kumar15, nakar19}.

Several lines of evidence of short gamma-ray bursts suggest that the central engine must be able to launch an ultra-relativistic outflow~\cite[e.g.,][]{ackermann10}, a constraint demanded by the compactness problem~\cite{ruderman75}. Constraints on the total energy requires that the outflow is collimated into a narrow jet~\cite[e.g.,][]{nakar07}, while rapid variability of the prompt emission implies the engine must be a compact object~\cite[e.g.,][]{nakar07,berger14}. Immediately, these constraints imply two potential central engines, a black hole or a neutron star. Moreover, one of the implicit requirements imposed by an ultra-relativistic outflow is having a relatively baryon-free environment where the jet is launched~\citep[e.g.,][]{nakar07,ciolfi18_sgrb}, such that the ultra-relativistic jet can launch efficiently, break out of the environment and stay relativistic. It is this constraint of having a clean environment, sometimes refereed to as the baryon-loading problem~\citep{shemi90},  that has led to debate about whether long-lived neutron stars can launch jets and power short gamma-ray bursts.

Assuming the central engine is a black hole, there are two main mechanisms for generating an ultra-relativistic jet that can produce a short gamma-ray burst: neutrino-antineutrino annihilation along the black hole spin axis~\citep{eichler89, ruffer98}, and the Blandford-Znajek mechanism~\citep{blandford77}. Although both processes may be responsible for launching jets, the Blandford-Znajek mechanism is more favoured due to energetic constraints, with the neutrino-antineutrino annihilation unable to produce high energy gamma-ray bursts~\citep[e.g.,][]{kyutoku18, ciolfi18_sgrb}, although we note that neutrino luminosity on long timescales is far from certain. 
In the case of a long-lived neutron star the mechanism for launching the jet is far from certain, although it is clear that if a jet is launched, it is likely through magnetic processes that tap into the large rotational-energy reservoir~\citep[e.g.,][]{bucciantini12, ciolfi19} or a combination of a neutrino-annihilation and magnetic processes~\citeg{perego17_neutrino, fujibayashi17_neutrino}. However, the problem as we alluded to above is the baryon-load of the environment. We refer the interested reader to more detailed reviews of gamma-ray burst jet launching~\citep[e.g.,][]{ciolfi18_sgrb}.

As we discussed earlier in Sec~\ref{sec:short_EM}, neutrino radiation from the nascent neutron star unbinds a non-negligible (potentially $\gtrsim 10^{-2} M_{\odot}$) amount of matter~\citep[e.g.,][]{fernandez16, metzger17}. This additional ejecta pollutes the surrounding environment, particularly along the spin axis~\citep[e.g.,][]{ciolfi17}. This pollution means that even tapping into the entire $\approx 10^{52}$ erg of rotational energy, the maximum Lorentz factor of a jet that forms is $\mathcal{O}(10)$, an order of magnitude smaller than required to alleviate the compactness problem~\citep{nakar07, murguia14, murguia17}, and two orders of magnitude smaller than measurements from short gamma-ray bursts~\citep[e.g.,][]{ackermann10}. This led \citet{murguia17} to argue that the formation of an ultra-relativistic jet has to wait until the formation of a black hole. One may think this opens up the possibility of a supramassive neutron star producing a short gamma-ray burst, however detailed analysis by~\citet{margalit15_supramassive} showed that for a range of equations of state, accretion disk and therefore jet formation is unlikely after the collapse of a supramassive neutron star. Given our current constraints on the maximum mass of neutron stars, this suggests up to 80\% of binary neutron star mergers will not produce a short gamma-ray burst.

\citet{beniamini17_grbengine} showed that to efficiently launch an ultra-relativistic jet, the energy per baryon ($\eta \propto \dot{E}/\dot{M}$) at the base of the jet must exceed a critical threshold ($\eta \gtrsim 100$). Following this result, one can derive a critical timescale for when this threshold is exceeded~\cite{beniamini20_delay}. For $B \sim\unit[3\times10^{16}]{G}$ this timescale is $\approx\unit[0.2]{s}$, short enough to be consistent with the delay seen in GRB170817A. Alternatively, for more typical parameters, this timescale is $\gtrsim\unit[3]{s}$, inconsistent with the delay of GRB170817A. This suggests that typical magnetars are unable to produce short gamma-ray bursts with a delay consistent with GRB170817A, although it is important to emphasise that it is not necessary for all short gamma-ray bursts to have the same delay from merger, implying a timescale $\gtrsim\unit[3]{s}$ is not necessarily problematic. Furthermore, given the complicated physics at play for launching a jet, it is worth questioning how reliably these scalings can distinguish between timescales within an order of magnitude. More importantly however is the dependency of this timescale on the magnetic field. As discussed in earlier sections, numerical simulations do not resolve the magneto-rotational and Kelvin-Helmholtz instabilities, underestimating the amplification of the magnetic field. If the magnetic field amplification is higher this will naturally alleviate this problem with the timescale for efficiently launching ultra-relativistic jets.

Several numerical simulations have shown the formation of a relativistic outflow from a long-lived neutron star~\citeg{giacomazzo13, ciolfi17, ciolfi19, ciolfi20_250ms}. However, these outflows do not have high-enough Lorentz factors (i.e., $\Gamma\gtrsim100$) to produce gamma-ray emission, and therefore explain short gamma-ray burst observations. The corollary is that long-lived neutron stars are not viable candidates for producing short gamma-ray bursts. However, recent numerical simulations have changed this conclusion~\cite{mosta20}. Including detailed physics of neutrinos in their simulation, \citet{mosta20} showed that neutrinos emitted by the nascent neutron star predominantly around the polar region end up preventing baryon pollution, which are therefore likely to aid the formation of a relativistic jet. It is worth noting that the \citet{mosta20} simulations still do not have outflows with sufficiently high $\Gamma$, however they propose this could be because they are still not including full neutrino transport, whereby pair annihilation could still substantially boost $\Gamma$ to the relevant regime.


While theoretical support is slowly arriving, there is considerable, although perhaps subjective, observational evidence for long-lived neutron stars being viable engines of short gamma-ray bursts. The $1.74$ s delay between the gravitational-wave signal, GW170817 and gamma-ray burst, GRB170817A has been calculated to be dominated by the timescale for the relativistic jet to reach the $\gamma$-ray emitting radius~\citep{ren20_delay, beniamini20_delay}. Independently, kilonova observations have been suggested to require a magnetised wind from a hypermassive neutron star that survived for up to $1$ s~\citep{metzger18}. In combination, these two independent arguments suggest that the relativistic jet in GRB170817A is likely launched by the differentially rotating neutron star before it collapsed into a black hole. If a differentially rotating neutron star can launch such a jet then long-lived neutron stars which also go through this evolutionary phase must be able to launch a jet as well.

Another more tentative observational evidence for a neutron star engine is the recent detection of a luminous kilonova from short GRB200522A~\citep{fong20_kilonova}. The kilonova is significantly brighter than expected from r-process nucleosynthesis alone and the characteristic features of the emission suggest additional energy from a neutron star. If true, this neutron star is most definitely stable and given the kilonova is detected as a counterpart to a short gamma-ray burst, the neutron star engine must have launched the jet that powered this gamma-ray burst. We note that given there are large systematic uncertainties with kilonova modelling~\citep[e.g.,][]{zhu20} and there may be alternative explanations for this exceptionally bright kilonova. Moreover, Gemini observations of GRB200522A only find a weak counterpart, contrary to expectations if a magnetar was present~\citep{oconnor21}. 

Perhaps the best observational evidence for long-lived neutron stars being able to launch relativistic jets and power short gamma-ray bursts are x-ray afterglow observations. The observations of a plateau~\citep[e.g.,][]{fan06, dallosso11, rowlinson13, lu15}, a sharp drop in luminosity~\citep[e.g.,][]{rowlinson10, rowlinson13, sarin20} and late-time x-ray flares~\citep[e.g.,][]{fan06} are all best interpreted by invoking the spin-down energy of a long-lived neutron star~\citep{zhang01, lasky17}. If long-lived neutron stars are powering the x-ray afterglows they must by definition also be responsible for launching the ultra-relativistic jet and producing prompt emission. 

Although these observational lines of evidence are not definitive, they do strongly suggest long-lived neutron stars can launch jets that produce short gamma-ray bursts, with more recent numerical simulations~\citep[e.g.,][]{mosta20} backing up this claim.

\subsection{Supramassive or stable}
Back-of-the-envelope calculations suggest supramassive stars born from neutron-star mergers are expected to survive for anywhere between $\sim\unit[10]{s}$ to $\sim\unit[4\times10^4]{s}$~\cite{ravi14}, which is broadly consistent with observationally-inferred collapse times from short gamma-ray burst x-ray afterglows~\cite{rowlinson13}.  However, a detailed look at the expected and observed distributions show the stars tend to collapse on shorter timescales than one might expect~\cite{fan13,ravi14,sarin20}.  Suggestions for the resolution of this discrepancy include excess gravitational-wave emission at early times~\cite{fan13,gao16,lu17,linLu19}, or deconfined quark matter changing the star's moment of inertia from that with normal matter~\cite{li16,li17,drago18}.  Hierarchical Bayesian inference of the population of collapse times for 18 supramassive neutron stars identified by the  $72$ \textit{Swift} x-ray afterglow observations suggest the story may be some combination of these two effects~\cite{sarin20}.  

These constraints have important consequences. Firstly, if these nascent neutron stars are really composed of deconfined quarks, then it suggests there is a temperature dependent hadron-quark phase transition.  Understanding where in the nuclear phase diagram such a transition occurs is critical to our understanding of the behaviour of nuclear matter~\cite[e.g.,][]{bauswein19, chatziioannou19}. An unexplored consequence of this phase transition is on the kilonova; additional neutrino flux from the cooling of a hybrid star cf. a purely hadronic star may mean that the electron fraction in the accretion disk outflow is higher than expected. In turn, this could make it difficult to infer the time the remnant collapsed based on the r-process elements synthesised~\cite[e.g., see][]{lippuner17, kyohei20_diversity}. We discuss the cooling of nascent neutron stars in more detail in Sec.~\ref{sec:thermal_evolution}. 

The implication that supramassive neutron stars are spinning down predominantly through gravitational waves has important consequences for the dynamics of these neutron stars, a point we discuss in more detail in later sections. 
More immediately relevant is the implication significant gravitational-wave spin down has on the electromagnetic observations. As discussed in Sec~\ref{sec:170817fate}, energetic constraints on GW170817 from kilonova~\cite[e.g.,][]{margalit17} and radio~\cite[e.g.,][]{ricci20_radio, schroeder20} observations put strong limits on the total energy of the ejecta. It has been proposed that this energy constraint indirectly rules out a long-lived neutron star in GW170817~\cite{margalit17}. However, if the supramassive star spins down predominantly through gravitational-wave emission and therefore collapses earlier than expected, then these constraints are weakened significantly. This also implies that future analyses utilising energy constraints to infer the nature of the remnant need to carefully consider energy losses through gravitational waves. However, it is worth stressing we have an incomplete, quantitative understanding of how much energy is radiated in gravitational waves. 

Finally, it is worth mentioning that the hierarchical analysis of collapse times~\cite{sarin20} predict a large fraction of supramassive stars collapse in $\lesssim\unit[100]{s}$. This is significant, because this is also approximately the time it takes \textit{Swift} to slew, implying a number of post-merger remnants may be observationally misidentified as being hypermassive because \textit{Swift} is not able to slew in time to see the result of the energy injection from the central engine.

\subsection{Thermal evolution}\label{sec:thermal_evolution}
The thermal evolution and history of a newly born neutron star has important consequences on several aspects of its evolution, including, for example, neutrino emission, the role of viscosity in dynamical and secular instabilities, the freezing out of the crust, and the transformation of the core to neutron superfluidity and proton superconductivity.

At very early times the star cools through neutrino emission~\citeg{shapiro83, perego19}. In general, the neutrinos do not free stream, implying the cooling timescale is set by the time it takes for the neutrinos to diffuse out of the star~\cite[e.g.,][]{rosswog03}. It is these neutrinos that affect the electron fraction of the ejecta, and hence alter the colour of the kilonova (see Sec.~\ref{sec:short_EM}).  An approximate cooling timescale given by the neutrino diffusion timescale is~\cite{paschalidis12}
\begin{align}
    t_{\rm cool}\approx 1200\left(\frac{M}{\unit[2.8]{M_\odot}}\right)\left(\frac{R}{\unit[15]{km}}\right)\left(\frac{E_\nu}{\unit[10]{MeV}}\right)^2{\rm ms},
\end{align}
where $E_\nu=\unit[10]{MeV}$ is the root-mean-square value of the neutrino energy found in simulations~\cite{rosswog03}.  

During the cooling of the remnant, the first major structural change is the formation of the crystalline lattice crust. This typically takes $\mathcal{O}({\rm days})$ for the base of the crust to begin forming, with a complete crust not forming for up to a year following birth~\cite[e.g.,][and references therein]{kruger15}.  In the stellar core, neutron superfluidity and proton superconductivity are expected at temperatures $\lesssim\unit[(5-9)\times10^8]{K}$ and $\lesssim\unit[(2-3)\times10^8]{K}$, respectively~\cite[see][and references therein]{page11,shternin11}.  While the development for superfluidity and superconductivity are certainly relevant for late-time evolution of neutron stars, e.g., in understanding glitches in young radio pulsars, it is not clear what observational impact it has on the evolution of post-merger remnants.

\subsection{Dynamical evolution}\label{sec:long_dynamics}
Nascent neutron stars born in binary neutron star mergers are differentially rotating with large poloidal magnetic fields in the range of $\sim 10^{14-16}$ G. This differential rotation winds up a toroidal component of the magnetic field roughly symmetrical to the rotation axis. One expects this toroidal component $B_t$ to dominate over the poloidal component $B_p$ deforming the neutron star into a prolate ellipsoid~\citep[e.g.,][]{cutler02, lander18}.
Misalignment between the rotation and magnetic axes result in precession~\citep[e.g.,][]{dallosso18, lander18}.
Internal dissipative processes drive the magnetic axis towards orthogonality with the spin axis~\citep{mestel72, jones76, cutler02}, minimising the energy of the system and also making the system an optimal emitter of gravitational waves \citep[e.g.,][]{cutler02, lasky16, dallosso18}. 
This instability is known as the ``spin-flip instability".

In reality, the evolution of the misalignment angle $\chi$ between the star's magnetic and rotation axes is significantly more complicated. 
In general, for a star deformed by a dominantly toroidal field, viscous dissipation increases $\chi$. Conversely, $\chi$ decreases for a star deformed predominantly through a poloidal field.
Early efforts to model the evolution of $\chi$ focused on determining the effects of this viscous dissipation.
However, the evolution of $\chi$ is in fact a complex mixture of dissipation, neutrino cooling, and the effects of the external torque from spin-down which also acts to decrease $\chi$~\citep{lander18}. 

The coupling between the spin-down and viscous effects proves to be critical in large parts of the neutron star parameter space and determining whether $\chi\rightarrow0^\circ$ (i.e., an aligned rotator that will not emit gravitational waves) or $\chi \rightarrow 90^\circ$ (i.e., an orthogonal rotator, an optimal emitter of gravitational waves)~\citep{lander18}.
This coupling is ignored in~\citet{dallosso18} by assuming that the timescale for spin flip is much faster than the spin down. This may not be true, and depends on the dissipation timescale, which is inversely proportional to the size of the deformation~\citep{ipser91}. The dissipation timescale is closely related to the spin-flip time-scale, an approximation of which is~\cite[e.g.,][]{lasky16},
\begin{equation}
    \tau_{\text{sf}} \approx \unit[8]{s} \left(\frac{R}{\unit[15]{km}}\right)^2\left(\frac{\left<\rho\right>}{\unit[10^{15}]{g cm^{-3}}}\right)\left(\frac{\epsilon_B}{10^{-3}}\right)^{-1} \left(\frac{\epsilon_\Omega}{0.3}\right)^{-2}\left(\frac{\xi}{\unit[10^{30}]{g\,cm^{-1}\,s^{-1}}}\right)^{-1}
\end{equation}
Here, $\left<\rho\right>$ is the volume-averaged density, $R$ is the neutron star radius, $\epsilon_B$ is the magnetic-field induced ellipticity, $\epsilon_{\Omega}$ is the rotational ellipticity, and $\xi$ is the bulk viscosity coefficient. 
For these parameters, it is difficult to imagine the spin-flip timescale being much faster than the spin-down timescale especially considering there are short gamma-ray bursts with observations on significantly shorter timescales where the emission is potentially derived from spin-down.
However, this timescale is extremely sensitive to the temperature $\xi \sim T^6$ and therefore the cooling history of the neutron star, which is not well understood.

The evolution of $\chi$ is critical for developing an understanding of what mechanism is responsible for radiating away the rotational energy of the long-lived neutron star~\citep[e.g.,][]{margalit18_grbslsn, lander20}. In the context of magnetars born in long gamma-ray bursts, \citet{margalit18_grbslsn} showed that an aligned rotator will almost exclusively power a gamma-ray burst, while a misaligned rotator will deposit some fraction of its rotational energy onto the supernova. Similar behaviour is to be expected in a long-lived neutron star born in a binary neutron star merger. In the first several seconds while $\chi\rightarrow0^\circ$ the rotational energy will be lost in powering the gamma-ray burst~\citep{lander20}. However once the system begins to orthogonalise i.e., $\chi\rightarrow 90^\circ$ the energy will be lost in gravitational waves and in powering the kilonova.

For neutron stars born in long gamma-ray bursts, the evolution of the braking index through coupling of the spin and magnetic axes has potentially been measured directly~\citep{mus19, cikintoglu20}. Such a model has not yet been fit to short gamma-ray burst observations. However, measurements of braking indices $n \lesssim 3$ of long-lived neutron stars from short gamma-ray bursts hint towards the evolution of $\chi$ for a number of objects~\citep{lasky17, sarin20radiative}, although it is worth noting there are other ways to explain such measurements~\citep{lasky17}.
If $\chi$ is evolving, it will be important to understand the timescales and the long-term evolution. This will provide important clues into the evolution of these objects into ordinary magnetars we see in our galaxy today. While more immediately, it will allow for more informed inferences about the radiation mechanisms that are tapping into the large rotational-energy reservoir of long-lived neutron stars. 

\subsection{Electromagnetic observations}\label{sec:long_obs}
In Sec~\ref{sec:jetlaunch} we discussed whether a neutron star could launch an ultra-relativistic jet and produce a short gamma-ray burst. Here, we discuss the broader electromagnetic imprints of a long-lived neutron star remnant. 

The longest numerical simulations of binary neutron stars and their remnants last approximately \unit[100]{ms} post merger~\citep[e.g.,][]{depietri18, ciolfi19, depietri20}, significantly shorter than the time where they could be used to provide insight into the electromagnetic signature of a long-lived neutron star. Such insights therefore rely predominantly on analytical and semi-analytic models~\citep[e.g.,][]{zhang01, dallosso11, metzger_piro14}.

The diverse predictions of electromagnetic signatures from long-lived post-merger remnants can all be primarily attributed to the large reservoir of rotational energy of the long-lived neutron star. Unlike hypermassive neutron stars which can trap a significant amount of their spin-down energy as they collapse into black holes~\citep[e.g.,][]{metzger17,shibata19}, supramassive and infinitely stable neutron stars will radiate a large fraction of this energy away with several electromagnetic consequences.

One of the signatures of spin-down energy is on the kilonova itself~\citep[e.g.,][]{yu13}. The injection of spin-down energy is either via a Poynting flux from a collimated jet~\citep{bucciantini12} or photons generated from dissipation of a magnetar wind~\citep{thompson04, zhang13}.
This increased energy creates a distinct impact on the kilonova lightcurve, altering the peak time and duration, while drastically increasing the luminosity of the kilonova compared to a kilonova only powered by radioactive decay. This latter characteristic implies the colour of the kilonova becomes bluer, simply due to higher luminosity translating to a higher effective temperature for a similar photospheric radius. Such effects are identical to the differences seen between ordinary and magnetar-driven supernovae~\citep{kasen10, nicholl20_slsn, margalit18_grbslsn}. As mentioned in Sec.~\ref{sec:170817fate}, this is one of the interpretations of the kilonova following GW170817~\citep{yu18}. The recent identification of an exceptionally bright kilonova following short GRB200522A~\citep{fong20_kilonova} provides further tantalising evidence for this magnetar-driven kilonova scenario. An additional imprint of this increased ejecta energy will be seen on the synchrotron radio signal generated from the interaction of the ejecta with the interstellar medium~\citep{hotokezaka15, horesh16, ricci20_radio, schroeder20}. Such constraints have been used to infer the fate of several gamma-ray bursts, suggesting that $<50\%$ of binary neutron star mergers make a long-lived neutron star~\citep{schroeder20}. However, we note that this may be overly constraining as they ignore rotational energy lost through other channels.

Even disregarding the impact of the spin-down luminosity, a long-lived remnant will show a distinctive characteristic feature on the kilonova. As discussed in Secs.~\ref{sec:short_EM} and~\ref{sec:thermal_evolution}, the nascent neutron star cools rapidly through neutrino cooling. This additional neutrino flux dramatically increases the electron fraction of the ejecta and suppresses the production of lanthanides~\citep[e.g.,][]{lippuner17,metzger17, kyohei20_diversity}.
Such ejecta would naturally be less opaque than more lanthanide-rich ejecta making the resulting kilonova dominantly ``blue" regardless of how much rotational energy is deposited into the kilonova ejecta.

Potentially some of the best electromagnetic observations of nascent neutron stars is the evidence of energy injection in the x-ray afterglows of gamma-ray bursts. A large fraction of x-ray afterglows have plateaus followed by sharp drops in luminosity~\citep[e.g.,][]{rowlinson13}, which are challenging to explain with the canonical model for afterglows that model the interaction of a relativistic jet with the surrounding environment. The spin-down energy from a nascent neutron star can naturally explain both these features~\citep[e.g.,][]{zhang01,rowlinson13}. However, the exact mechanism that extracts this rotational energy is still uncertain. Either the spin-down energy is extracted directly from the remnant itself~\citep[e.g.,][]{zhang01,rowlinson13, lasky17, strang19} with a constant efficiency, or indirectly through energy injection at the afterglow shock interface~\citep[e.g.,][]{cohen99,dallosso11, sarin20radiative}. Systematic model selection suggests the latter model, with a generalised braking index better explaining the observations~\citep{sarin20radiative}, at least for the small subset (eight) of short and long gamma-ray bursts analysed. Such a model also self-consistently explains x-ray flares that are seen at the onset of the plateau phase, providing additional support for this model.

An alternative way to generate x-ray emission with a magnetar is through the interaction of a magnetar wind with the merger ejecta~\citep[e.g.,][]{yu13, metzger_piro14, siegel16, strang19, strang20}. Assuming the magnetar is completely enshrouded in the merger ejecta, the magnetar wind energy will be dissipated via shocks or magnetic re-connection creating a hot nebula behind the ejecta. This hot nebula will be comprised of photons and electron/positron pairs analogous to a pulsar wind nebula~\citep[e.g.,][]{metzger_piro14}. Initially, most of the spin-down energy will be lost in expanding the nebula and ejecta along with it. This expansion will eventually reduce the optical depth allowing photons at various wavelengths to diffuse out once the diffusion timescale becomes shorter than the expansion timescale~\citep{metzger_piro14}. 
While such a model explains several observational features of x-ray afterglows, it is problematic as this model cannot explain x-ray emission at early-times ($T\lesssim\unit[10]{hr}$) as the ejecta is still optically thick. In this model, early-time x-ray observations must be from synchrotron emission from the interaction of the relativistic jet with the surrounding environment, with the neutron star component becoming dominant after the ejecta becomes optically thin. Such a constraint is, however, problematic, given sharp drops in x-ray plateaus are seen as early as \unit[100]{s} after the prompt~\citep[e.g.,][]{sarin20}. There are two ways to reconcile this issue. First, the timescale for the ejecta to become optically thin need not be as long as \unit[10]{hr}, as this timescale is dependent on the opacity of $r$-process elements, which are far from certain~\citep[e.g.,][]{tanaka20, barnes20}. Second, and perhaps more importantly, the simple assumption that a magnetar is wholly enshrouded in the merger ejecta may not be valid, with potential holes due to piercing by the ultra-relativistic jet that produced the gamma-ray burst or by Rayleigh-Taylor instabilities in the ejecta itself~\citep[e.g.,][]{strang19, strang20}. 

An important question to consider with the spin-down energy is whether the radiation is emitted isotropically or beamed in a particular direction, perhaps along with the jet that produces the gamma-ray burst prompt emission itself. The answer to this question could shed critical insight into the emission mechanism. Observations and interpretations of CDF-S XT2~\citep{xue19} as a magnetar seen off-axis suggests the magnetar spin-down energy is emitted isotropically. However, isotropic emission potentially leads to violations of the total energy budget for certain gamma-ray bursts believed to be powered by nascent neutron stars~\citep[e.g.,][]{beniamini17}. By contrast, if the spin-down energy is collimated, it must be through a mechanism that can be sustained for long timescales to explain the late-time x-ray observations. One explanation for the collimation originates from the interaction of the magnetar wind and the surrounding environment. Magnetohydrodynamic simulations suggest that the deceleration of the magnetar wind due to the dense environment of the merger ejecta may collimate the wind into a jet for sufficiently high spin-down luminosity~\citep{bucciantini12}.
However, such a mechanism seems unfeasible, as the jet will become unstable and susceptible to magnetohydrodynamic instabilities once the spin-down luminosity of the long-lived remnant drops~\citep{porth13, metzger_piro14}.

Ultimately, testing detailed models with better observations will allow us to determine what physical processes are relevant. Beyond these more immediate electromagnetic observations, long-lived neutron stars may also be responsible for fast radio bursts, to which we now turn our attention.

\subsection{Fast radio bursts}
Fast radio bursts are millisecond duration pulses of coherent radio emission.  In general they come from cosmological distances, with dispersion measures significantly larger than galactic values.  We refer the reader to Refs.~\cite{cordes19, zhang20_frbreview} for recent reviews on fast radio bursts. 

Ever since the original discovery~\citep{lorimer07}, subsequent identification of fast-radio burst repeaters~\citep{spitler14,spitler16} and localisation of its host galaxy~\citep{chatterjee17,tendulkar17}, nascent magnetars have been invoked to explain these enigmatic astrophysical phenomena~\citep[e.g.,][]{popov13,lyubarsky14, kulkarni14, katz16,beloborodov17, metzger17_frb, metzger19_frb, katz18, lu_kumar18, margalit19_frb_bns, lu20_frb}. While it is possible that there are two distinct populations of fast radio bursts, characterised as repeaters and isolated bursts~\citep[e.g.,][]{falcke14,zhang17_frbcomb,zhang20_frb}, a non-repeating population may be disfavoured on rate-based arguments~\citep[e.g.,][]{nicholl17_frb, ravi19}. 
We note there are numerous other alternative progenitor models, in fact, as recently as 2019, theoretical models outnumbered the number of events themselves~\citep{platts19}.
However, this all changed with the recent watershed discovery of a fast radio burst from a galactic magnetar, SGR1935+2154, which provided the smoking-gun observation for the magnetar origin for fast radio bursts.

SGR1935+2154 is a galactic soft gamma repeater first identified by \textit{Swift} as a potential gamma-ray burst candidate~\citep{stamatikos14}, it is associated with a supernova remnant at a distance $d\approx6$ kpc~\citep{gaensler14,zhou20}, with an estimated surface magnetic field $B\approx 10^{14}$ G and age (based on the supernova remnant association) $\gtrsim 16$ kyr~\citep{zhou20}. On April 28, 2020, a millisecond duration radio pulse was independently detected from this source by CHIME~\citep{chime20_sgr} and STARE2~\citep{stare2_sgr} in coincidence with a bright X-ray burst~\citep{zhang20_sgr}. 
Various analyses confirmed the analogous nature of the coherent radio emission with cosmological fast radio bursts albeit with significantly lower energy (at least $\sim 25$ times) than typical cosmological fast radio bursts. 
This observation provided unequivocal evidence that magnetars are the progenitors of at least some fast radio bursts~\citep[e.g.,][]{margalit20_sgr}, with the discrepancy in energies attributed to the old age of this magnetar, weaker magnetic field, and slower rotation~\citep[e.g.,][]{lu20_frb, beloborodov20}. 

Observations of SGR1935+2154 provide evidence that young, rapidly rotating magnetars produce some fast radio bursts. However, it is unclear whether these magnetars are ones born in core-collapse supernovae or in binary neutron star mergers. The first repeating fast radio burst FRB121102 was localised to a low metallicity dwarf star-forming galaxy~\citep{chatterjee17, tendulkar17} and also associated with a persistent radio source~\citep{marcote17}. The host galaxy properties and the persistent radio source are best interpreted as the emission from a young magnetar embedded in the expanding supernova ejecta~\citep[e.g.,][]{omand18, margalit18_121102} pointing towards a superluminous supernovae/long gamma-ray burst origin for FRB121102. The Australian Square Kilometer Array Pathfinder (ASKAP) has since localised another fast radio burst FRB180924~\citep{bannister19}, finding the host galaxy properties and offsets to be comparable to short gamma-ray bursts.
FRB180924, therefore, provided the first possible evidence for a binary neutron star remnant merger origin for fast radio bursts. Since then, more fast radio bursts have been localised in host galaxies strongly suggestive of binary neutron star merger remnant origins~\citep[e.g.,][]{ravi_frb190523}. Rate-based arguments suggest all fast radio bursts are repeaters~\citep[e.g.,][]{ravi19}, implicitly demanding long-lived sources, i.e., long-lived neutron star remnants. 
Moreover, as supramassive neutron stars have been shown to collapse on relatively short timescales~\citep[e.g.,][]{sarin20} where the kilonova ejecta is still optically thick, implying fast radio bursts would not escape to an external observer~\citep[e.g.,][]{margalit19_frb_bns}. The timescale for this ejecta to become optically thin ranges from weeks to months, implying any fast radio bursts associated with binary neutron star mergers must be from infinitely stable neutron stars~\citep[e.g.,][]{margalit19_frb_bns}. 

While the progenitor model of fast radio bursts can be confidently confirmed as a nascent magnetar, the exact mechanism that generates the coherent radio emission is unknown. The various models can be broadly divided into two categories: those for which the emission is close to magnetar i.e., a pulsar-like mechanism involving the magnetosphere~\citep[e.g.,][]{katz16,lu20_frb}, and those for which the emission mechanism involves relativistic shocks similar to gamma-ray bursts, where the coherent radio emission is generated far from the magnetar~\citep[e.g.,][]{beloborodov17,metzger19_frb}. Future observations of fast radio bursts will be able to shed light into which mechanism is correct.  

\subsection{Gravitational-wave emission and detection}
\subsubsection{Gravitational-wave emission}
Long-lived neutron stars also emit gravitational waves, but we have an incomplete understanding of which mechanisms are relevant, how long they are active for, or how much energy is emitted. As we discuss later, this uncertainty weakens our ability to detect gravitational waves from a long-lived neutron star while also weakening our ability to infer the fate from indirect energetic constraints from electromagnetic observations~\cite[e.g.,][]{shibata19}.

Three main instabilities are relevant for producing gravitational waves in long-lived post-merger remnants. These are the spin-flip, bar-mode and \textit{r}-mode instabilities~\citep[e.g.,][]{cutler02,lai95,shapiro00,shibata00, andersson01,andersson03}. Note there are potentially other mechanisms, such as gravitational-wave emission due to the formation of mountains from fall-back accretion~\citep{sur20}. Here we briefly discuss these gravitational-wave emission mechanisms following on from the discussion in Sec~\ref{sec:long_dynamics}. For a detailed review of gravitational waves from neutron stars we refer the reader to~\citep[e.g.,][and references therein]{lasky15_review}.

The precessional spin-flip instability drives a nascent neutron star to become an orthogonal rotator and therefore an optimal emitter of gravitational waves~\citep{cutler02}. Past studies~\citep[e.g.,][]{lander18} have shown that once a system evolves to being near-aligned or near-orthogonal, it does not further evolve.  However, more recent work~\citep[e.g.,][]{lander20} has shown that including the effect of a neutrino-driven wind can change the late-time behaviour, with orthogonal rotators slowly decreasing the angle between magnetic and rotation axes over hundreds of years to become more aligned. \citet{lander18} also showed that nascent neutron stars would become orthogonal rotators for sufficiently large toroidal magnetic fields $B_t \gtrsim 10^{14}$G. Such magnetic fields are naturally expected in binary neutron star mergers~\citep[e.g.,][]{giacomazzo13, kiuchi14,ciolfi19,mosta20}, making it likely that long-lived neutron stars born in binary neutron star mergers become orthogonal rotators and stay that way for a long time.

As discussed in Sec~\ref{sec:long_dynamics}, the timescale for orthogonalisation is far from certain, but is a critical component for building gravitational-wave waveform models for the spin-flip instability.
An aligned rotator does not emit gravitational waves. As the angle between the magnetic and rotation axes increases, the star will emit gravitational waves dominantly at the spin frequency, or twice the spin frequency, with a host of other potential values that depend on the precession timescale~\citep[e.g.,][]{jones02,lasky13}. Building gravitational-wave waveform models for this evolution requires careful modelling of the orthogonlisation timescale, including the evolution of the misalignment angle between the rotation and magnetic axes. 
If the orthogonlisation timescale is short compared to the overall gravitational-wave emission timescale, then the uncertainty associated with the evolution of the misalignment angle can be largely ignored.  Such an assumption would be detrimental to detection prospects if the orthogonalisation timescale is comparable to the timescale for gravitational-wave emission.

Assuming orthogonalisation has already taken place,~\citet{sarin18} built a model for the gravitational-wave signature of long-lived post-merger remnants. Unfortunately, detection prospects are not good with second-generation gravitational-wave interferometers; for detection at greater than a few Mpc, the amount of emitted gravitational-wave energy must exceed the rotational energy budget of the system.

The critical quantities determining the energy produced in gravitational waves for gravitational-wave dominated spin down (i.e., the most optimistic scenario cf. when the spin down is dominated by electromagnetic torques) are the initial spin period and ellipticity. The strong magnetic fields expected in a binary neutron star merger deform the nascent neutron star, with the size of this deformation dependent on the magnetic field and the internal geometry of the field. For simple stellar models, the ellipticity can be approximated as~\cite{cutler02}
\begin{equation}
\epsilon_{B} \approx 10^{-6}\left(\frac{\left\langle B_{\mathrm{t}}\right\rangle}{10^{15} \mathrm{G}}\right)^{2},
\end{equation}
where $\langle B_{\mathrm{t}}\rangle$ is the volume-averaged toroidal magnetic field strength. Observational inferences based on the collapse times of supramassive neutron stars from short gamma-ray bursts suggests $\epsilon_B \sim 10^{-3}$~\citep{gao16}. Such large ellipticities require $\langle B_{\mathrm{t}}\rangle \sim 10^{16}$--$\unit[10^{17}]{G}$, i.e., toroidal fields that are 1-2 orders of magnitude stronger than inferred values of the poloidal field.  It is an open question whether such magnetic fields are dynamically stable~\citep[e.g.,][]{braithwaite09,lasky11,ciolfi12,akgun13,herbrik17,sur20}.

Another mechanism to generate gravitational waves in a newly-born neutron star is through unstable $f$ modes, also known as the bar-mode instability. This comes in two varieties: dynamical and secular. To activate the dynamical instability, the ratio of a star's rotational kinetic energy $T$ to gravitational binding energy $W$ must be $T/|W| \geq 0.24$~\citep[][although note this number is dependant on the unknown equation of state]{lai95,shapiro00,baiotti07, corsi09}, while the secular instability is active for $T/|W| \geq 0.14$. For realistic equations of state, the dynamical instability is only activated when the star is differentially rotating~\citep[e.g.,][]{friedman78,corsi09,ravi14,lasky16}. The growth of the dynamical instability is therefore halted when the magnetic field damps differential rotation, which is governed by the Alfv\'en timescale of $\mathcal{O}(1)$s~\citep{shapiro00}. Therefore, the dynamical bar mode is only expected to be active in the first seconds of a neutron star's life.

More relevant for long-term gravitational-wave emission is the secular instability. It is worth noting that several equations of state that support masses $\gtrsim\unit[2]{M_{\odot}}$ do not have rigidly-rotating equilibrium solutions with $T/|W|\geq0.14$, implying the secular bar-mode instability cannot be active for those equations of state~\cite{ravi14}.  However, if active, the secular bar mode grows to large non-linear amplitudes on timescales of $\mathcal{O}(10$--$100)$\,s~\citep{doneva15}. For the instability to grow, the nascent neutron star has to cool to a temperature below $\sim\unit[10^{10}]{K}$, such that bulk viscosity does not suppress the instability. As we have discussed in Sec~\ref{sec:thermal_evolution}, a newly-born neutron star is cooled to this temperature a few seconds after the merger, making the secular-bar mode a potentially-dominant mechanism for gravitational-wave emission~\cite{corsi09}. 

Another oscillation mode that will likely lead to the gravitational-wave radiation are the $r$ modes; low-frequency toroidal oscillations for which the Coriolis force is the restoring force. These oscillations are retrograde in the co-moving frame and prograde in the inertial frame,\citep{andersson01, andersson03}, making them always unstable to the Chandrasekhar-Friedman-Schutz (CFS) instability~\citep{chandrasekhar70, friedman78}. Whether this instability is active is dependent on a delicate balance between gravitational-wave radiation which drives up the size of the instability, and viscous processes that dampen the oscillations. For nascent neutron stars, the dominant viscous force is bulk viscosity, which in turn depends sensitively on neutron star microphysics, cooling history, and rotation~\citep[e.g.,][]{andersson03,lasky15_review}.

Recently, an interpretation of the x-ray afterglow of GRB090510 suggests that the observations support gravitational-wave losses due to $r$-mode oscillations through the measurement of the braking index $n=7$~\citep{linLu19}. This interpretation is contentious, most notably because a braking index of $n=7$ is not necessarily a reliable indicator of spin-down through $r$-mode gravitational-wave emission~\citep[e.g.,][]{alford14,alford15}

Long-lived neutron stars could also emit gravitational waves from mountains formed through fall-back accretion~\cite[e.g.,][]{piro12, melatos14, sur20}. However, given the relatively little amount of material ejecta in a binary neutron star merger, this mechanism is likely only relevant for neutron stars born in supernovae and long gamma-ray bursts.

All the different mechanisms described above suggest it is quite likely that long-lived neutron stars spin down, at least in some part, through gravitational wave emission. This is to an extent observationally verified by the observed collapse time distribution~\citep{sarin20}, which suggests $\sim 70\%$ of these neutron stars are spinning down predominantly through gravitational-wave radiation. However, knowing which mechanism is active and for how long can only be verified through the direct detection of gravitational waves from these objects, something to which we now turn.

\subsubsection{Gravitational-wave detection methods}
Searching for gravitational waves from long-lived neutron stars suffers from some of the same problems as traditional searches for continuous gravitational waves~\citep[see][and references therein]{riles13}. The long-duration signals expected implies traditional matched-filtering approaches are computationally unfeasible~\citep[e.g.,][]{brady98} requiring semi-coherent methods that are less sensitive than fully coherent methods. Similarly, the uncertainty in the gravitational-wave modelling necessitates the use of unmodelled searches, which by design are not as sensitive as modelled searches. 

The LIGO/Virgo Collaborations searched for gravitational waves from a potential long-lived post-merger remnant from GW170817 on intermediate~\cite[$\lesssim 500$ s;][]{abbott17_gw170817_postmerger} and long~\cite[$\lesssim8$ d;][]{abbott19_gw170817_postmergerII} timescales (the latter timescale being set by the length of data available following the merger).  No viable candidate was found, however upper limits on the gravitational-wave strain and energy were derived.

A number of complementary pipelines were used in the LIGO/Virgo searches.  The robustly-named Stochastic Transient Analysis Multi-detector Pipeline~\cite[STAMP;][]{thrane11} looks for tracks of excess power in cross-correlated data from the two detectors using seedless clustering algorithms~\cite{thrane13,thrane15}. Coherent Wave Burst~\cite[cWB;][]{klimenko16} was used with a similar setup as for the short-duration search; see Sec.~\ref{sec:short_GW}. The Hidden Markov Tracking method using the Viterbi algorithm~\cite{suvorova16,sun19} searches for quasichromatic signals with unknown frequency evolution and stochastic timing noise. The Adaptive Transient Hough~\cite{krishnan04,oliver19} assumes the signal's frequency evolution can be modelled as a power law, with the amplitude and frequency described by the generalised millisecond magnetar model~\cite{lasky17,sarin18}.  Finally, the Generalized FrequencyHough algorithm~\cite{antonucci08, astone14, miller18} is a pattern-recognition technique that maps time-frequency points to lines in frequency-spin down space.

Four of the aforementioned algorithms are unmodelled searches, and one is modelled.  Upper limits for the unmodelled searches were derived using two theoretically-motivated signal models; the generalised millisecond magnetar model that describes a spinning-down neutron star with arbitrary braking index~\cite{lasky17,sarin18}, and a waveform model~\cite{corsi09} where the star's spin evolution is dictated by the gravitational-wave driven secular Chandrasekhar-Friedman-Schutz instability~\cite{chandrasekhar70, friedman78}.  

The derived upper limits from the searches do not necessarily bode well for future detections of gravitational waves from long-lived neutron star remnants. Throughout the entire explored parameter space, the distance at which a source could have been observed from the remnant of GW170817 was, at best, just $\sim\unit[1]{Mpc}$, cf. the actual distance of $\sim\unit[40]{Mpc}$.  

Although aLIGO/Virgo's sensitivity is predicted to improve by a factor of a few over the sensitivity at the time GW170817 was detected, the prospect for detection with second-generation telescopes is still grim. From an observational perspective, aLIGO/Virgo could eventually be sensitive to mergers at best at $\sim\unit[10]{Mpc}$.  Theoretical estimates are consistent with this~(e.g., see Refs.~\cite{corsi09,fan13,dallosso15, doneva15}, although see~Refs.~\cite{dallosso18,lasky16,sarin18} for more pessimistic estimates).
However, in practice, the most optimistic of these estimates may require nonphysical quantities of gravitational-wave energy, for example in excess of the total rotational energy budget of the system~\cite{sarin18}.  A successful detection of gravitational waves from a long-lived post-merger remnant may therefore have to wait until A+~\cite{Aplus}, or even third-generation detectors such as Einstein Telescope~\cite{EinsteinTelescope} or Cosmic Explorer~\cite{CosmicExplorer}.

\section{Conclusions and Outlook}
The era of gravitational-wave and electromagnetic multi-messenger astronomy began spectacularly with GW170817~\cite{abbott17_gw170817_detection, abbott17_gw170817_multimessenger,abbott17_gw170817_gwgrb}, a watershed event that provided invaluable insights across several domains of high energy astrophysics. Observations of GW190425 transformed our understanding of the formation of binary neutron stars, highlighting the flaw in the simple assumption that the mass distribution of cosmological binary neutron stars follows the same mass distribution as ones observed locally in our galaxy~\citep[e.g.,][]{abbott20_gw190425}. 
The future of electromagnetic and gravitational-wave multi-messenger astronomy looks bright, with upgrades to both gravitational-wave and electromagnetic detectors in progress~\cite[e.g.,][]{abbott16_LRR,ztf19} and proposed for the future~\cite[e.g.,][]{vera_rubin, EinsteinTelescope,CosmicExplorer, OzHF20}. 

Notwithstanding our limited understanding of the maximum mass of neutron stars and the binary neutron star mass distribution, we expect potentially up to $\sim 80 \%$ of neutron star mergers to produce some form of neutron-star remnant~\cite{margalit19}. However, at least for the near future, the smoking gun observation of gravitational waves from short or long-lived remnants is unlikely. As such, the fate of the remnant will need to be inferred through the suite of electromagnetic observations. 

In the near future, electromagnetic observations, in particular, the early x-ray afterglow and kilonova observations provide the best probe into the nature of the remnant. Late time radio limits on the energy of the ejecta will also provide clues into the nature of the remnant. The latter analyses will require a good understanding of the different channels of energy emission, in particular gravitational waves. If future research concludes that long-lived remnants cannot launch ultra-relativistic jets capable of producing short gamma-ray bursts. This would imply that short gamma-ray bursts are a biased and relatively small fraction of binary neutron star mergers, something that will become telling as the rate of binary neutron star mergers becomes better constrained. Furthermore, any discrepancy in binary neutron star merger and gamma-ray burst rates must therefore be explained by neutron star-black hole mergers, a statement which has significant implications on the properties of such binaries. 

Although x-ray afterglow and kilonova observations are the best electromagnetic channels for determining the fate of the post-merger remnant, they alone may also not be definitive without further development and testing of models, requiring input from theorists, simulations and observers. In particular, the idea of using the colour of kilonova to probe the lifetime of the remnant is fraught with difficulties such as viewing angle dependence~\cite[e.g.,][]{darbha20}, and the uncertain impact of the jet-ejecta interaction~\cite[e.g.,][]{nativi20}. These uncertainties, combined with our incomplete knowledge of nuclear reactions and opacities of r-process elements, implies significant systematic uncertainties for inferring the properties of kilonovae from observations~\cite[e.g.,][]{zhu20}.

Inferring the fate of the remnant from the early-time x-ray afterglow observations may be more reliable, particularly if the emission from the long-lived neutron star is isotropic (as suggested by the observations of CDF-S XT2~\cite[e.g.,][]{xue19}). This is different from the emission expected from the interaction of the jet with the interstellar medium (the physics known to be responsible for the afterglow in normal circumstances), which is strongly affected by relativistic beaming~\cite[e.g.,][]{totani02,granot02}. However, models for the emission from nascent neutron stars are in their infancy, with significant development required such that they accurately reflect all the critical physics~\cite[e.g.,][]{metzger_piro14,lasky17,mus19, strang19,sarin20radiative}. Jet structure could also explain such observations without requiring a neutron star remnant; in such a scenario, systematic model selection may provide the answer~\cite[see e.g.,][]{sarin19}. 

The observations of an x-ray plateau with a sharp drop in luminosity following a binary neutron star merger may provide the most definitive electromagnetic evidence for the fate of a binary neutron star merger. Here again there may be different emission mechanisms such as a reverse shock formed from the interaction of a relativistic jet with the surrounding interstellar medium~\cite[e.g.,][]{vaneerten18, lamb19_reverse}, or radial stratification of the jet such that it is refreshed at late times~\cite[e.g.,][]{lamb20_refreshed}. High-latitude emission~\cite[e.g.,][]{oganesyan20, ascenzi20}, or fall back accretion~\cite{desai19} that may also explain the sharp drop in luminosity without requiring a supramassive neutron star. Fortunately, if the gamma-ray burst is observed off-axis, several of these scenarios become less likely to be the cause~\cite[e.g.,][]{vaneerten18}. However, to ensure this sharp drop in luminosity is observed, the electromagnetic counterpart of a binary neutron star merger must be identified quickly, on timescales as short as $100$s, which will be difficult in the near future. In light of these theoretical and observational issues, it is the combination of various electromagnetic phenomena and richer data confronted with more detailed models that can shed light into the nature of the remnant.

In this review, we have discussed the fate of binary neutron star mergers GW170817 and GW190425 and potential other neutron star mergers seen as short gamma-ray bursts. We have reviewed all possible outcomes of a binary neutron star merger from the prompt collapse into a black hole to the formation of an infinitely stable neutron star, discussing their observational signature, evolution, and prospects for gravitational-wave detection. 
As observations of binary neutron star mergers grow in number, understanding the fate of the remnant will become increasingly more important due to its far-reaching implications, such as on the nuclear equation of state, gamma-ray bursts, kilonovae, fast radio bursts, and beyond. At least in the near future, the lack of smoking-gun gravitational-wave observations means the fate must be inferred from electromagnetic observations. The promise of richer and more frequent observations confronted with better models ensures this will be an exciting endeavour.

\begin{acknowledgements}
We are grateful to Carl Knox for the creation of Fig.~\ref{fig:post_merger_evolution}. We also thank Andrew Melatos, Toni Font Eleonora Troja, and the anonymous reviewers for their helpful comments on the manuscript. 
This work is supported through Australian Research Council (ARC) Centre of Excellence CE170100004, ARC Future Fellowship FT160100112 and ARC Discovery Project DP180103155. N.S is supported by an Australian Government Research Training Program (RTP) Scholarship.
\end{acknowledgements}
\bibliographystyle{apsrev4-1} 
\bibliography{post_merger_review}

\end{document}